\documentclass{article}
\usepackage{scalerel}
\usepackage{adjustbox}
\usepackage{blindtext, rotating}
\usepackage{amsthm}
\usepackage{bbm}
\usepackage{amsmath}
\usepackage{amssymb}
\usepackage{natbib}
\usepackage{enumerate}
\usepackage{graphicx}
\usepackage{amssymb}
\usepackage{multirow}
\usepackage{verbatim}
\usepackage{subfigure,placeins}
\usepackage{array}
\newcolumntype{C}[1]{>{\centering\arraybackslash}p{#1}}

\usepackage{enumerate}
\usepackage{bbold}

\newcommand\R{\mathbb{R}}
\newcommand\E{\mathbb{E}}
\newcommand\opt{{\scaleto{\text{OPT}}{3pt}}}

\title{Causal inference under mis-specification: adjustment based on the propensity score}
\author{David A. Stephens$^1$ \and Widemberg S. Nobre$^2$ \and Erica E. M. Moodie$^3$ \and  Alexandra M. Schmidt$^3$ \\ \ \\
	$^1$Department of Mathematics and Statistics, McGill University, Canada\\
	$^2$Departamento de M\'etodos Estat\'isticos, Universidade Federal do Rio de Janeiro, Brazil\\
	$^3$Department of Epidemiology and Biostatistics, McGill University, Canada\\
	}

\date{\today}
	
\begin{document}
		
\maketitle

		\begin{abstract}
			We study Bayesian approaches to causal inference via propensity score regression. Much of the Bayesian literature on propensity score methods have relied on approaches that cannot be viewed as fully Bayesian in the context of conventional `likelihood times prior' posterior inference; in addition, most methods rely on parametric and distributional assumptions, and presumed correct specification.  We emphasize that causal inference is typically carried out in settings of mis-specification, and develop strategies for fully Bayesian inference that reflect this.  We focus on methods based on decision-theoretic arguments, and show how inference based on loss-minimization can give valid and fully Bayesian inference.  We propose a computational approach to inference based on the Bayesian bootstrap which has good Bayesian and frequentist properties. 

		\end{abstract}

	\section{Introduction}\label{sec:intro}

In the study of the causal relationship between an exposure (or treatment) and an outcome, bias in the estimation of the exposure effect may occur due to confounding if the exposure is not an experimental intervention.  Confounding exists whenever the exposure assignment is dependent on predictors that also influence the outcome.  If the dependence of outcome on exposure and predictors is modelled correctly, standard regression is adequate to obtain correct inference about the exposure effect.   When correct specification cannot be guaranteed, the \textit{propensity score} can be used to break the dependence between confounders and exposure, to create \textit{balance} in the distribution of confounders across exposure groups, and facilitate correct inference.  This paper studies how the propensity score can be deployed in a Bayesian causal analysis.

Adjustment via the propensity score can be carried out using regression, inverse weighting, stratification or matching. In regression settings, parametric models are proposed to represent the propensity score and the (expected) outcome given the propensity score. In frequentist approaches, adjustment is carried out by estimating parameters in the propensity score and the outcome models separately. In a fully Bayesian framework, such a two-step analysis is uncommon; it would be more natural to fit a single \textit{joint} model for the treatment and outcome.   This has led to discussion as to how Bayesian methods can be used in the causal setting, and even whether Bayesian methods are valid. There is a growing literature on sophisticated procedures for performing Bayesian causal analysis, but in a fully Bayesian framework, some aspects of the methodologies deployed appear non-standard and not justified via Bayesian logic.

We address these issues in this paper. Section \ref{sec:Background} recaps the regression approach to causal estimation, and section \ref{sec:Exchangeability} describes how the key to valid Bayesian causal inference results from the assumption of exchangeability of the observable quantities to be modelled, which can be derived through de Finetti's representation, and a review of Bayesian adjustments using the propensity score. Section \ref{sec:Bdti} describes Bayesian decision-theoretic inference which gives the framework for inference under mis-specification, and section \ref{sec:BBootsec} gives the non-parametric computational strategy that we deploy.  Section \ref{sec:FormalFrame} recasts the conventional Bayesian approach in the decision-theoretic framework.  We provide simulation studies in section \ref{SimStudy}, more complicated inference settings in section \ref{sec:Extensions}, and conclude with a discussion in section \ref{conc}.

We note here that Bayesian methods that do not rely on the propensity score are also quite widely used: these methods utilize flexible parametric or non-parametric procedures to represent the outcome model as a function of the treatment and other predictors and attempt to avoid mis-specification.   These methods are certainly useful, and the inferential theory supporting such one-stage analyses is more straightforward.  However, such flexible outcome regression models cannot estimate the causal effect of interest in all cases, such as those where a more general target of inference is defined.  These models are not the primary focus of this paper.  Similarly, we will not discuss Bayesian matching methods in detail, although some comments are given in section \ref{conc}.



\section{Background} \label{sec:Background}

To formulate causal inference estimation, \textit{potential} or \textit{counterfactual outcomes} are often used.  Potential outcomes, $\{Y(z)\}$ for $z$ in some putative treatment set, represent the outcomes that would be observed if treatment level $Z$ was set to $z$. If exposure $Z$ takes two levels labelled $\{0,1\}$,  the potential outcomes represent values of outcome $Y$ that would be observed had exposure been set by intervention to $z=0,1$ respectively \citep{Neyman1923,rubin1974,holland1986statistics}.  We consider $n$ subjects, and for the $i$th subject, let $Y_i$ be the outcome of interest, $Z_i$ be the exposure, and ${X}_i = (X_{i1},X_{i2},\ldots,X_{ip})^\top $ be a $p$-dimensional vector of confounders.  We denote the data observation space $\mathcal{X}$.

\subsection{The average treatment effect (ATE)}

Under an assumption of no unmeasured confounding or ignorability, so that $\{Y(z)\} \perp Z | X$,  the \textit{\textit{Average Treatment Effect}} (ATE) for a binary treatment is defined by
	\begin{equation}
	\tau = \E[Y(1)] - \E[Y(0)].
	\label{eq:potentialate}
	\end{equation}
If $Z$ is assigned \textit{independently} of covariates $X$, $Z \perp X$, the ATE, $\tau$, is defined as
	\begin{equation}
	\tau = \E_{Y|Z}[Y|Z=1] - \E_{Y|Z}[Y|Z=0] = \E_X[\E_{Y|X,Z}[Y|X,Z=1] - \E_{Y|X,Z}[Y|X,Z=0]].
	\label{eq:ate}
	\end{equation}
This definition differs notationally from the formulation via \textit{counterfactual} or \textit{potential outcomes} \citep{rubin1974} or the \textit{do-operator} \citep{pearl2009causality}, but under the independence assumption is equivalent.  Equation (\ref{eq:ate}) defines a marginal (over $X$) estimand, although conditional (subset-specific) estimands may also be defined. The calculation in \eqref{eq:ate} can be mimicked in the observed data to yield the estimate
\begin{equation}\label{eq:ate-est}
\widehat \tau = \frac{1}{n} \sum_{i=1}^n (\E_{Y|X,Z}[Y|X=x_i,Z=1] - \E_{Y|X,Z}[Y|X=x_i,Z=0])
\end{equation}
but this requires knowledge of the conditional expectation $\E_{Y|X,Z}[Y|X=x,Z=z]$ for all $(x,z)$.  Typically, this expectation would be represented using a regression model, and if this model is misspecified, incorrect inference about $\tau$ in the presence of confounding, when the independence assumption does not hold, and $X$ is also associated with $Y$.

\subsection{The role of the propensity score}
\label{sec:role}

To estimate the ATE in the presence of confounding, \cite{rosenbaum1983central} showed that if the exposure assignment is ignorable and $b(X)$ is a \textit{balancing score}, defined so that $X \perp Z | b(X)$, the ATE can be evaluated by averaging analyses carried out conditional on $b(X)$.  If $Z$ is binary, a typical choice for the balancing score is the propensity score, where $b(X) = \Pr[Z=1|X]$.  Conditioning on the propensity score allows estimation of $\tau$ in the presence of confounding when the conditional model for $Y$ given $X$ and $Z$ is not correctly specified by breaking the dependence between $X$ and $Z$,

Propensity score regression represents the expected outcome conditioned on the exposure, confounders and propensity score. The ATE $\tau$ from (\ref{eq:ate}) can be evaluated as
	\begin{equation}
	\tau = \E_{X}\left\{\E_{Y|X,B,Z}[Y|X,b(X),Z=1] - \E_{Y|X,B,Z}[Y|X,b(X),Z=0]\right\}
	\label{atePS}
	\end{equation}
as $b(X)$ is a balancing score, based on a model for $\E_{Y|X,B,Z}[Y|X=x_i,B=b(x_i),Z=z]$ for a modified version of \eqref{eq:ate-est} -- see for example \cite{rosenbaum1983central}.

Typically $b(\ldotp)$ is represented using a parametric model, $b(x) \equiv b(x;\gamma)$, with $\gamma$ estimated from the observed $Z$ and $X$ data.  However, the balancing result $X \perp Z \ | \ B$ only holds when $b(X;\gamma)$ correctly characterizes the probability that $Z=1$ for any given $X$;  this corresponds to the existence of a true value $\gamma_0$ of $\gamma$ which defines the function precisely.   For $\gamma \neq \gamma_0$, the method of proof of \cite{rosenbaum1983central} does not work to establish balance; see Appendix section \ref{sec:AppPS} for a summary of the argument.  Therefore, in a correctly specified parametric formulation of the propensity score, to yield balance, we \textbf{must} identify a single point in the parameter space, and use that to define the propensity score.  If $\gamma_0$ is not known, we must resort to substituting a consistent estimator $\widehat \gamma$ for $\gamma_0$, and then the required balancing result will hold asymptotically.

\subsection{An illustrative model}
\label{sec:im}
Suppose the observed outcome data are generated according to the structural model
\begin{equation}\label{eq:RMN0}
Y_i = X_{0i} \xi + Z_i \tau + \epsilon_i
\end{equation}
where for $p$-dimensional parameter $\xi$ the term $X_{0i} \xi$ defines the true treatment-free mean model, and $\tau$ defines the ATE.  If a regression model matching this specification is fitted using least squares, then the resulting estimator for $\tau$ is consistent.  Similarly, if $Z$ is assigned independently of $X$, then the estimator for $\tau$ is consistent even if the treatment-free mean model is mis-specified.  However, if the model is mis-specified and $Z$ and $X$ are not independent, then the estimator of $\tau$ is in general inconsistent due to confounding.  As demonstrated by \cite{RobinsMarkNewey1992} the regression model
\begin{equation}\label{eq:RMN1}
Y_i = b(X_i) \phi + Z_i \tau + \epsilon_i,
\end{equation}
where $\phi$ is scalar parameter, yields a consistent estimator of $\tau$, albeit one whose variance is at least as large as the variance of the estimator arising from the correctly specified model. An `augmented' model that contains an additional `prognostic' linear predictor term $ X_i \beta$ involving nuisance parameter $\beta$, that is, with
\begin{equation}\label{eq:RMN2}
Y_i = X_i \beta + b(X_i) \phi + Z_i \tau + \epsilon_i
\end{equation}
can be fitted as an attempt to reduce the variance for $\widehat \tau$; note, however, that the inclusion of this augmenting term is not necessary for consistent estimation of $\tau$ provided the propensity score model is correctly specified. Least squares then provides a semiparametric estimation approach.

If a parametric model $b(x) = b(x;\gamma)$ is used, then parameter $\gamma$ must be consistently estimated for the adequate adjustment.   A plug-in estimation procedure, where $\gamma$ is replaced by $\widehat \gamma$, and the regression utilizes $b(x;\widehat \gamma)$ is typically used and corresponds to the `feasible' E-estimator of \cite{RobinsMarkNewey1992}.  It is justified in part by the asymptotic independence of $\widehat \gamma$ and $(\widehat \phi, \widehat \tau)$ \citep{henmieguchi2004}.  The extended model \eqref{eq:RMN2} has the advantage of additional inferential robustness: if the $X_i \beta$ component is correctly specified (i.e.~reflects the data-generating mechanism), the estimator of $\tau$ will be consistent even if the propensity score is not correctly specified. This is known as \emph{double robustness}.  If the data generating structural model contains a more general treatment effect structure, the propensity score regression approach must be modified.   For example, if the model takes the form
\begin{equation}\label{eq:RMN00}
Y_i = X_{0i} \xi + Z_i M_{0i} \psi + \epsilon_i
\end{equation}
where $\psi$ is a $q \times 1$ vector parameter and $M_{0i}$ is a $1 \times q$ vector of predictors, the ATE is $\E[M_{0i}] \psi$.  This quantity (and the parameters $\psi$) can be consistently estimated using the propensity score regression approach based the model
\begin{equation}\label{eq:RMN20}
Y_i = X_i \beta + b(X_i) M_{0i} \phi + Z_i M_{0i} \psi + \epsilon_i
\end{equation}
where now $\phi$ is a $q \times 1$ parameter, that is, with an interaction term involving the propensity score configured to match the treatment effect model.  This construction is necessary to ensure that confounding via the open paths that involve the interaction terms is also removed by conditioning on the propensity score.  Further modifications are necessary if the structural model is extended beyond the linear; see section \ref{sec:Extensions}.

\section{Bayesian inference under exchangeability} \label{sec:Exchangeability}

The key construction for any Bayesian inference problem to be solved under an assumption of exchangeability of the observable quantities is de Finetti's representation, which leads to the standard definitions of likelihood, prior, parameters and the notion of `correct specification'. If $\{O_i\}_{i=1}^\infty$ is a sequence of exchangeable observable quantities, where each $O_i$ takes values on $\mathcal{X}$, the de Finetti representation of the joint density of any collection of size $n \geq 1$ of the observables is
\begin{equation}\label{eq:deforig}
p_{O}(o_{1:n}) = \int \prod_{i=1}^{n} f_{O}(o_i;\theta) \pi_0(\theta) d \theta
\end{equation}
where $\pi_0(\theta)$ is the prior distribution on parameter $\theta$ presumed to take values in parameter space $\Theta$. Bayesian inference about $\theta$ is made via the posterior distribution
\begin{equation}\label{eq:postorig}
\pi_n(\theta) = \frac{\prod\limits_{i=1}^{n} f_{O}(o_i;\theta) \pi_0(\theta)}{p_{O}(o_{1:n})} .
\end{equation}
If $o_{1:n}=(x_{1:n},y_{1:n},z_{1:n})$ $\theta$ characterize the joint distribution.  We have
	\begin{equation}\label{eq:factorization0}
	p_{O}(o_{1:n}) = p_X(x_{1:n})p_{Z|X}(z_{1:n}|x_{1:n})p_{Y|Z,X}(y_{1:n}|x_{1:n},z_{1:n}).
\end{equation}
Decomposing $\theta = (\eta, \gamma, \zeta)$, and assuming independent prior structure, we require that the three components in (\ref{eq:factorization0}) each admit a de Finetti representation based on what we term \textit{conditional} exchangeability assumptions \citep{saarela2022role}.  For $n \geq 1$, the triples $(X_i,Y_i,Z_i)$, $i=1,\ldots , n$, are assumed to be conditionally independent given $\theta$, and the Bayesian specification is completed after defining a probability distribution $\pi_0(\theta) \equiv \pi_{0}(\eta)\pi_{0}(\gamma)\pi_{0}(\zeta)$. Specifically
	\begin{align}\label{eq:factorization01}
	p_X({ {x}_{1:n}}) &= \int \prod_{i=1}^{n}f_X({x}_i;\zeta) \pi_0(\zeta)d\zeta, \nonumber\\
	p_{Z|X}({z_{1:n}| {x}_{1:n}}) &= \int \prod_{i=1}^{n}f_{Z|X}(z_i|x_i;\gamma) \pi_0(\gamma) d\gamma, \\
	p_{Y|X,Z}({y_{1:n}| {x}_{1:n},z_{1:n}}) &= \int \prod_{i=1}^{n}f_{Y|X,Z}(y_i|{x}_i,z_i;\eta) \pi_0(\eta)d\eta . \nonumber
	\end{align}
This formulation proposes that in the data generating model the $Y_i$s are conditionally independent given the $(X_i,Z_i)$ pairs and parameter $\eta$.  This is a standard assumption in the frequentist parametric sequel, and would hold in any conventional regression model.  The full probability model for observables and unobservables can be decomposed as
	\begin{equation}
f_{X}(x_{1:n};\zeta)f_{Z|X}(z_{1:n}|x_{1:n};\gamma)f_{Y|Z,X}(y_{1:n}|x_{1:n},z_{1:n};\eta)\pi_{0}(\zeta)\pi_{0}(\gamma)\pi_{0}(\eta)
	\label{jointModel}
	\end{equation}
with the usual conditional independence decompositions of the `likelihood' terms.  The prior independence assumption is natural in light of the conditional exchangeability formulation in \eqref{eq:factorization0}.  This leads to the posterior distribution $\pi_n(\eta,\gamma,\zeta)$ in the usual way. Under standard assumptions, the posterior distribution converges as $n \longrightarrow \infty$ to a unique degenerate limit at a single point $(\eta_0,\gamma_0,\zeta_0)$, and the data generating model is in fact factorized $f_X({x};\zeta_0) f_{Z|X}(z|{x};\gamma_0) f_{Y|X,Z}(y|{x},z;\eta_0)$.  The Bayesian model is considered correctly specified if this limiting behaviour holds.

The formulation above is parametric, but extensions to the non-parametric case where $\theta$ is infinite dimensional are straightforward. We regard a valid Bayesian approach as one which relies on the de Finetti representation (or equivalent) for observable quantities in the data generating model, with inference following a decision-theoretic argument, as outlined in section \ref{sec:Bdti}.  Note that under exchangeability, the de Finetti representation defines (up to the choice of the prior) the complete probabilistic specification for the model, whether or not we opt to depend on it for inference.  Furthermore, it determines the frequentist characteristics of Bayesian inference procedures.

\subsection{Existing approaches to Bayesian causal inference}
\label{sec:Bm}

A parametric Bayesian analysis based on the true model \eqref{eq:RMN0} or proposed model \eqref{eq:RMN1} would proceed in a standard fashion. The marginal posterior distributions for $\tau$ derived from \eqref{eq:RMN0} and \eqref{eq:RMN1} are in general different.  However, model \eqref{eq:RMN0} is essentially an `oracle' model to which we do not have access.  In this case, it is relatively straightforward to show that as $n$ increases, the posterior distribution for $\tau$ derived from \eqref{eq:RMN1} becomes concentrated at the true (data generating) value of the ATE present in the structural model, despite the mis-specification present in \eqref{eq:RMN1}.  This asymptotic calculation hypothesizes an increasingly large sample of data drawn from the same probability model.  These arguments hold for the extended model \eqref{eq:RMN2}.

If $b(x)$ is replaced by $b(x;\gamma)$ in \eqref{eq:RMN1} or \eqref{eq:RMN2}, and $\gamma$ is treated as an unknown parameter, the question arises whether this log-likelihood, coupled with the log-likelihood for $\gamma$ itself, should be used as the basis of a three-parameter posterior in the parameters $(\phi,\tau,\gamma)$.  It is not evident on first inspection whether this posterior, or the bivariate posterior based on $(\phi,\tau)$ for some plug-in value $\widehat \gamma$ as in the frequentist approach is justified in a formal Bayesian inference setting.  \cite{zigler2016central} summarizes the most commonly used approaches, and describes directions in which the Bayesian formulation may be developed productively.  We summarize some of the key elements below.

\smallskip

\noindent \textbf{Joint Bayesian modelling:} The Bayesian propensity score model proposed by \cite{mccandless2009bayesian} assumes a joint parametric model and for \eqref{eq:RMN2}, the joint model considers conditional models $f_{Z|X}(z|x) = f_{Z|X}(z|x;\gamma)$ and $f_{Y|X,Z,B}(y|x,z,b(x)) = f_{Y|X,Z,B}(y|x,z,b(x;\gamma),\beta)$.  This leads to a joint likelihood function for $(\gamma,\beta,\phi,\tau)$:
	\begin{equation}
	\mathcal{L}_n(\beta,\phi,\tau,\gamma) = \prod_{i=1}^{n} f_{Z|X}(z_i|{x}_i; \gamma)f_{Y|X,Z}(y_i|z_i,{x}_i,b({x}_i;\gamma);\beta,\phi,\tau),
	\label{eq:jointLikelihood}
	\end{equation}
with inference carried out using Markov chain Monte Carlo (MCMC) -- specifically the Gibbs sampler -- by sampling recursively from the two full conditional distributions $\pi (\gamma|\beta,\tau,y_{1:n},z_{1:n},\mathrm{x}_{1:n})$ and $\pi (\beta,\phi,\tau|\gamma,y_{1:n},z_{1:n},\mathrm{x}_{1:n})$, along with any additional parameters that appear in the proposed models.

\smallskip

\noindent \textbf{Cutting feedback:} The joint model based on \eqref{eq:jointLikelihood} does not create the required balance, or correct appropriately for confounding, due the presence of what is termed \textit{feedback}, and the marginal posterior for $\tau$ does not concentrate at the true value. To overcome this, \cite{mccandless2010cutting} proposed that the full conditional distribution of $\gamma$ should be independent from the rightmost term of the likelihood in equation \eqref{eq:jointLikelihood}; \begin{equation}\label{eq:gammaposterior}
\pi_n(\gamma) \propto f_{Z|X}(z_{1:n}|\mathrm{x}_{1:n};\gamma)\pi_0(\gamma).
\end{equation}
This is known as the \textit{cutting feedback} approach which can be implemented as follows: a sample of size $L$ of $\pi_n(\gamma)$ is produced, and then used to construct propensity score sampled values $ {b}^{(l)}_{i} =\Pr[Z_i=1|X_i=x_i;\gamma^{(l)}]$, where $\gamma^{(l)}$ denotes the $l$-th sample from $\pi_n(\gamma)$. Then, a sample of size $L$ is obtained for the outcome parameters, with the $l$-th sample, for $l=1,\cdots ,L$ being generated from
\begin{equation}\label{eq:gammaposteriorCF}
\pi_n^{(l)}(\beta,\phi,\tau) \propto f_{Y|X,Z,B}(y_{1:n}|\mathrm{x}_{1:n},z_{1:n},{b}^{(l)}_{1:n};\beta,\phi,\tau)\pi_0(\beta,\phi,\tau).
\end{equation}

\smallskip

\noindent \textbf{Two-step inference:} A \textit{two-step} procedure \citep{zigler2013model} assumes complete separation between the exposure and outcome models.  First, a point estimate of $\gamma$ is obtained from $\pi_n(\gamma)$ computed via \eqref{eq:gammaposterior}. This point estimate is then used to construct an estimate of the propensity score, $\widehat{b}_{i} =f_{Z|X}(1|x;\widehat{\gamma})$, which is then plugged into the outcome model. A posterior sample is then obtained from
\begin{equation}\label{eq:gammaposterior2S}
\pi_n(\beta,\phi,\tau) \propto f_{Y|X,Z,B}(y_{1:n}|z_{1:n},\mathrm{x}_{1:n},\widehat{b}_{1:n};\beta,\phi,\tau)\pi_0(\beta,\phi,\tau).
\end{equation}

\smallskip
		
In the cutting feedback and two-step approaches, it is not immediately clear how the inferential uncertainty concerning $\gamma$ in the estimation of $\tau$ should be handled.  Several methods to evaluate the variance of the posterior distribution of $\tau$ have been proposed; see for example \cite{kaplan2012two}. The cutting feedback approach attempts to account for the uncertainty in the estimation of $\gamma$ by direct sampling from $\pi_n(\gamma)$ in \eqref{eq:gammaposterior} with posterior computation for the remaining parameters being carried out conditionally on each sampled value of $\gamma$; the two-step approach as described above ignores the uncertainty in $\gamma$, but an adjustment based on Taylor expansions around $\widehat \gamma$ can be implemented \citep{GMS2016}.

\subsection{Current literature}
\label{sec:Literature}

It is not universally accepted that fully Bayesian inference is possible using the fitted propensity score in a regression as in \cite{RobinsMarkNewey1992}, or via other methods such as inverse probability weighting (see \ref{sec:ipw}), as such methods involve a plug-in strategy is not fully Bayesian; see the discussion of \cite{saarela2015bayesian}.  For example, it is contended that if the propensity model is unknown and must be estimated, the plug-in estimation of $b(x)$ is contrary to conventional Bayesian inference based on a `likelihood times prior' formulation.  This issue can be resolved using more general Bayesian decision-theoretic logic, and a Bayesian analysis under model mis-specification.

Despite such objections, there has been a marked increase in research on Bayesian methods for causal quantities based on propensity score adjustment \citep[see, for example,][]{adhikari2019nonparametric,comment2019survivor,Geneletti2019,Samartsidis2020,Nethery2020,Liu2020}. While sharing a common goal of adjusting for bias due to confounding with a Bayesian lens, it is clear that consensus has not been reached on how to perform inference with propensity score-based approaches. For instance, \citet{comment2019survivor}, \citet{Nethery2020}, and \citet{Liao2020} all use an approach that succeeds in cutting feedback, using the propensity score to create a matched sample; these authors view the matching step as part of a `design' rather than analytic phase of the analysis. \cite{bornn2019moment} use a form of joint modelling of the treatment and outcome, as do \citet{ray2018semiparametric}.   Two-step approaches are widely used, although there is no agreement in the literature on whether to plug in fixed quantities (such as a posterior mean or mode) or random (draws from the posterior). For instance, \citet{Vegetabile2020} use a Bayesian non-parametric approach to estimate the propensity score which is then plugged into a standard (frequentist) estimator of the average treatment effect. \citet{wang2019bayesian} use propensity score regression, conditioning on the expected value of the propensity score. In contrast, \citet{xu2018bayesian} take a propensity regression approach to estimate the quantile (rather than average) treatment effect, conditioning on draws from the posterior distribution of the propensity score. \citet{hahn2020bayesian} sample the estimated propensity score's posterior distribution, incorporating the samples  into a nonlinear regression model for the outcome (including heterogeneous treatment effects) using additive regression trees. \citet{Liu2020} use inverse weighting in a two-step procedure and propagate uncertainty using the Bayesian bootstrap; see also \citet{GMS2016}. Other authors have combined aspects of Bayesian and frequentist modelling to address complex models.  \citet{davis2019addressing} use approximate Bayesian methods to estimate both a propensity score and an outcome model, and then combine predictions from these into a frequentist doubly-robust estimator in a spatial modelling context.  \citet{Antonelli2020HighDim} consider the high-dimensional case, also using Bayesian methods to estimate both a propensity score and an outcome model and computing a doubly-robust estimator by averaging over draws from the posterior distribution of the parameters of these models.

Models \eqref{eq:RMN1} or \eqref{eq:RMN2} are simple compared to some of the approaches described above, but serve to illustrate the relevant theoretical issues.  Flexible models that attempt to model the outcome directly can be extremely useful in capturing the causal relationship by overcoming issues of mis-specification.  Similar models are also widely used to represent the treatment-confounder relationship in a flexible model for the propensity score and, despite some drawbacks, such models can be effective.  The methods described in this paper are relevant to any form of propensity score modelling.

\section{Bayesian decision-theoretic inference}

\label{sec:Bdti}

The Bayes estimate is a function of the observed data that minimizes the Bayes risk, or the posterior expected loss for some loss function $\ell(t,\theta): \Theta \times \Theta \longrightarrow \R^+ $, that is
\[
\widehat \theta = \arg \min_{t \in \Theta} \E_{\pi_n}[\ell(t,\theta)] = \arg \min_{t \in \Theta} \int \ell(t,\theta) \pi_n(\theta) \ d \theta.
\]
If the loss function can be written
\begin{equation}\label{eq:IntegralLossSingle}
\ell(t,\theta) = \int u(s,t) f_O(s;\theta) \ d s  = \E_{f_O}[u(S,t);\theta]
\end{equation}
for some function $u(s,t): \mathcal{X} \times \Theta \longrightarrow \R^+$, then the estimation problem can be rewritten
\begin{equation}\label{eq:PostPredEst}
\widehat \theta = \arg \min_{t \in \Theta} \displaystyle \int u(s,t) \left\{\displaystyle \int f_O(s;\theta) \pi_n(\theta) \ d \theta \right\} ds = \arg \min_{t \in \Theta} \E_{p_n}[u(S,t)]
\end{equation}
where $p_n(s)$ is the posterior predictive distribution implied by the Bayesian specification.  For example, if, for $t \in \Theta$, $u(s,t) = - \log f_O(s;t)$, (see \cite{bernardo1979expected}) we have that
\begin{align}\label{eq:PostPredEstLog}
\widehat \theta & = \arg \max_{t \in \Theta} \displaystyle \int \left\{ \displaystyle \int \log f_O(s;t) f_O(s;\theta) \ ds \right\} \pi_n(\theta) \ d \theta.
\end{align}
For example, in the Normal model with $f_O(s;t) \equiv Normal(t,1)$, the calculation becomes
\[
\arg \min_{t \in \Theta} \displaystyle \displaystyle \iint (s-t)^2 \phi(s-\theta) \ ds \pi_n(\theta) \ d \theta
= \displaystyle \int \left\{ \displaystyle \int s \phi(s-\theta) \ ds \right\} \pi_n(\theta) \ d \theta = \int \theta \pi_n(\theta) \ d \theta
\]
where $\phi(.)$ is the standard Normal pdf, that is, the estimate is the posterior mean.   Equation \eqref{eq:PostPredEst} indicates that Bayesian parameter estimation can be formulated as a prediction problem if an appropriate loss function is defined.  Equation \eqref{eq:IntegralLossSingle} depends on an integral over a single variable $s$ that can be taken to be a single `future' variate drawn from $f_O(\ldotp;\theta)$, but the formulation extends to $m$ independent `future' variates, and can be expressed via the $m$-fold posterior predictive.

\subsection{The Gibbs posterior}

The standard Bayesian posterior distribution can be justified \citep{zhang2006a,JiangTanner2008,bissiri2016} as the solution to the variational expected loss minimization problem
      \begin{equation}
\label{eq:bobj}
\pi_n(\theta) = \arg\inf\limits_{\mu \in \mathcal{M}_{\pi_0}} \left\{ \int_{\Theta}\ell \left(o_{1:n},\theta\right)\mu\left(d \theta \right) + {\mathcal{K}\left(\mu,\pi_0\right)}\right\}
\end{equation}
where $\mathcal{M}_{\pi_0}$ is the space of probability measures that are absolutely continuous with respect to the prior (measure) $\pi_0$, $\mathcal{K}\left(\mu,\pi_0\right)$ is the Kullback-Leibler divergence between measure $\mu$ and $\pi_0$, and $\ell \left(o_{1:n},\theta\right) =- \sum_{i=1}^n \log f_O(o_i;\theta)$ is a loss function measuring the value of $o_{1:n}$ for learning about $\theta$ (see \citet{bernardo1979expected}).  It follows that
\begin{equation}\label{eq:GibbsProblem}
\pi_n(\theta) =\arg\inf\limits_{\mu \in \mathcal{M}_{\pi_0}} \int \log \left[\frac{ \mu\left(\theta\right)}{\exp\left(-\ell\left(o_{1:n},\theta\right)\right)\pi_0\left(\theta\right)}\right]\mu\left(d \theta\right),
\end{equation}
which yields the conventional posterior $\pi_n(\theta)$ by properties of the Kullback-Leibler divergence.   More generally, if the loss function is not specified as minus a log density, the solution to the loss minimization problem has been termed the \textit{Gibbs posterior}.   For the log-density specification for $\ell(\ldotp,\ldotp)$, this method is equivalent to the de Finetti formulation, but more general specifications are also possible.  Equation \eqref{eq:bobj} thus provides an alternative but also fully Bayesian decision-theoretic solution.

For bivariate data, the same logic can be applied.   Suppose that $o_i = (x_i,y_i)$, and the model is specified by two parameters $(\theta,\phi)$, with loss function
\begin{equation}\label{eq:bobjoint}
\ell \left(o_{1:n},(\theta,\phi) \right) = - \sum_{i=1}^n \log f_X(x_i;\theta) - \sum_{i=1}^n \log f_{Y|X}(y_i|x_i;\phi).
\end{equation}
The variational formulation \eqref{eq:GibbsProblem} leads to the joint posterior $\pi_n(\theta,\phi)$, and under an independent prior specification, the two parameters are \textit{a posteriori} independent.

\subsection{Bayesian inference under mis-specification}

\label{sec:Bims}

Broadly, mis-specification of a Bayesian model arises either if the `likelihood' model -- the conditional density of the observables given the parameters -- does not match $f_O$, or if the true value $\theta_0$ does not lie in the support of the prior.  In such cases, there is no guarantee of reliable statistical behaviour.  However, certain mis-specified models can have utility; for example, the model in \eqref{eq:RMN1} is not the data generating model, and yet can provide consistent frequentist inference provided the propensity score model is correctly specified.  In this section we examine some aspects of mis-specification.

Suppose initially we retain the data generating likelihood model $f_O(\ldotp;\theta_0)$, but consider the implications for inference in a second model with density $f$ having support $\mathcal{X}$, parameterized by $\vartheta \in \Theta^\prime$.  That is, while assuming the data are generated by $f_O$, we wish to perform inference for $\vartheta$ acknowledging that $f$ is mis-specified. Conventional Bayesian inference for $\vartheta$ can be performed using a likelihood based on $f$, but it is difficult to justify the resulting posterior as the focus of inference since the model is mis-specified; see, for example, \cite{walker2013} and its discussion.  The Bayesian decision theoretic framework can be deployed, however.  Define loss function $\ell(t^\prime,\theta) : \Theta^\prime \times \Theta \longrightarrow \R^+$ by
\[
\ell(t^\prime,\theta) = \mathcal{K}(f_O(\ldotp;\theta),f(\ldotp;t^\prime)) = \int \log \left(\frac{f_O(s;\theta)}{f(s;t^\prime)} \right) f_O(s;\theta) \ ds = \E_{f_O}[u_\theta(S,t^\prime);\theta]
\]
where $u_\theta(s,t^\prime) = \log \left(f_O(s;\theta)/f(s;t^\prime)\right)$, which extends the calculation in \eqref{eq:IntegralLossSingle} to allow the function $u(\ldotp,\ldotp)$ to depend on $\theta$ -- note that the resulting optimization over $t^\prime$ may still not depend on $\theta$.  By arguments equivalent to those leading to \eqref{eq:PostPredEstLog}, we have that
\begin{equation}\label{eq:PostPredEstLogMisspec}
\widehat \vartheta = \arg \max_{t^\prime \in \Theta^\prime} \displaystyle \int \left\{ \displaystyle \int \log f(s;t^\prime) f_O(s;\theta) \ ds \right\} \pi_n(\theta) \ d \theta.
\end{equation}
To compute the posterior distribution for $\vartheta$, we may use a simulation-based strategy; if a single sampled variate $\theta^{(l)}$ is generated from $\pi_n(\theta)$, then we may convert this into a sampled variate $\vartheta^{(l)}$ from the posterior for $\vartheta$ by performing the transformation
\begin{equation}\label{eq:PostPredEstLogMisspecSamp}
\vartheta^{(l)} = \arg \max_{t^\prime \in \Theta^\prime} \displaystyle \int \log f(s;t^\prime) f_O(s;\theta^{(l)}) \ ds
\end{equation}
and then replicate this for $l=1,\ldots,L$.  In each of the expressions, the integral with respect to $s$ may not be analytically tractable, but can be approximated using Monte Carlo by sampling $s_{k}, k=1,\ldots,N$ from $f_O(\ldotp;\theta)$, and computing
\[
\vartheta^{(l)} = \arg \max_{t^\prime \in \Theta^\prime} \sum_{k=1}^N \log f(s_k;t^\prime).
\]
Standard Bayesian theory is used to compute the posterior for $\theta$, and the posterior for $\vartheta$ is computed (using the relevant integral forms) via deterministic transformation.

The Kullback-Leibler loss can be modified to reflect quantitative statements about $\vartheta$ in the approximating model.  For example, we may specify
\begin{equation}\label{eq:uPrior}
u_\theta(s,t^\prime) = \log \left(\frac{f_O(s;\theta)}{f(s;t^\prime)} \right) + \log u_0(t^\prime)
\end{equation}
for some non-negative function $u_0(\ldotp)$ with domain $\Theta^\prime$ that does not depend on $\theta$ or $s$.  This additional term essentially functions as (minus) a log prior distribution on $\vartheta$, although as we explicitly acknowledge that the model $f$ is mis-specified, and $\vartheta$ has no real-world interpretation, this interpretation may be problematic for some Bayesians.  In any case, the maximizations leading to the estimate $\widehat \vartheta$ in \eqref{eq:PostPredEstLogMisspec} and sampled variate $\vartheta^{(l)}$ in \eqref{eq:PostPredEstLogMisspecSamp} can be modified accordingly.

\subsection{Conscious mis-specification and modularization}

The formulation of inference under mis-specification is inspired by the reasoning that inference concerning an approximating model may be of interest in its own right (for example, simplicity of interpretation). In addition, note that the calculation in \eqref{eq:PostPredEst} does not require explicit computation of the posterior $\pi_n(\theta)$, so in principle a representation of, approximation to, or samples drawn directly from $p_n(s)$ can be used to compute the estimate or posterior sample for $\vartheta$.  Such a strategy would be useful if complex models such as flexible Bayesian models or artificial neural networks were used to construct prediction techniques.  In the causal inference setting, the parameters of interest are not defined in the actual data generating model, but rather are quantities defined with respect to some hypothetical data generating process where confounding is not present.   It is possible to construct examples where even a correctly specified regression model, say, cannot yield consistent estimators of the causal effect of interest, although these examples typically need to have more complex structural forms than those in \eqref{eq:RMN0}, involving multiple treatments.  We return to these examples in section \ref{sec:Extensions}.

Such `conscious' mis-specification has direct relevance in the causal setting, but it has also been argued that similar calculations, where the data generating model does not correspond to the inference model, may be relevant in Bayesian calculations more generally.   \cite{bayarri2009} argue for a form of Bayesian inference based on `modularization' of the model, where a form of stagewise analysis in complex models is used. Motivated by formulations based on Bayesian mis-specification, \cite{jacob2017better} provide extensive evidence that such modularized inference can be advantageous in Bayesian settings, including a study of the empirical properties of propensity score regression estimators using the methods from section \ref{sec:Bm}.

\subsection{Connection to estimating equations}

\label{sec:EstEquation}

If the utility function in \eqref{eq:PostPredEst} or $u_\theta(s,t^\prime)$ is differentiable with respect to its second argument with derivative $\dot{u}_\theta(s,t^\prime)$, the optimization problem can be re-stated as a root-finding problem where we must solve
\begin{equation}\label{eq:EstEq}
\E_{p_n}[\dot{u}_\theta(S,\vartheta)] = 0,
\end{equation}
for $\vartheta$ to obtain the estimate or sampled variate as in the calculation described in section \ref{sec:Bims}.   In the Monte Carlo version, we sample $s_k,k=1,\ldots,N$ from the posterior predictive $p_n(\ldotp)$, and solve the (Bayesian) estimating equation
\begin{equation}\label{eq:EstEqMC}
\sum_{k=1}^N \dot{u}_\theta(s_k,\vartheta) = 0.
\end{equation}
Note that if the utility function is specified as minus a log density, then a scale or dispersion parameter, $\lambda$ say, may be present, but may be irrelevant to the estimation of $\vartheta$ in \eqref{eq:EstEqMC}. If $\lambda$ is estimated as a nuisance parameter, the utility optimization can typically be carried out for the parameters of interest and nuisance parameter separately in two sub-problems.  In this case, the parameters may still exhibit posterior dependence due their common dependence on the posterior predictive distribution or sampled values.  If the estimation of $\lambda$ is to be included, the utility function must chosen with some care; we might require that the (joint) utility is not monotonic in $\lambda$.

\noindent \textbf{Note:} It is tempting for a Bayesian analysis to mimic the frequentist approach to estimating equations, adopting a general form of \eqref{eq:EstEqMC} and performing root-finding to produce the estimate.  Again this approach needs careful implementation.  Consider for example the loss-based approach to defining a standard posterior as in \eqref{eq:bobj}; for a specified loss $\ell(\ldotp,\ldotp)$, the Gibbs posterior is automatically defined as being proportional to $\exp\{-\ell(o_{1:n},\theta)\} \pi_0(\theta)$.  For example, if the $\theta$ is scalar and $\ell(o,\theta) = |o - \theta|$, the procedure is immediately equivalent to using a double exponential likelihood model with known scale parameter.  This equivalence illustrates the potential for loss-based derivation of the posterior to be quite restrictive.  This cautionary note is also relevant to Bayesian estimation for mis-specified models described in this section: $u_\theta(s,\vartheta)$ must be a true, well-calibrated expression of the utility of specifying the approximating parameter as $\vartheta$ for generic datum $s$ when the data generating model is $f_O(\ldotp;\theta)$.

\section{Bayesian non-parametric formulation}
\label{sec:BBootsec}

In each case described in the previous section, the Bayesian model for the observables must not be mis-specified, and in general this is hard to guarantee using a parametric formulation. We now implement the ideas from section \ref{sec:Bdti} in the causal setting using a non-parametric model.

\subsection{The Dirichlet process model}

In order to weaken the parametric assumption concerning $f_O$, we allow $\theta$ to become an infinite dimensional parameter describing the distribution of $O$.  Suppose that $F_O(.)$ parameterizes unknown distribution function of the data with true value $F_0$, such that in reality $O_1,\ldots,O_n \sim F_0(.)$ are independent; this interpretation is consistent with the de Finetti formulation, with the $F_0(.)$ interpreted as the limiting empirical cdf derived from the exchangeable sequence.  The Dirichlet process model $DP(\alpha, G)$ is a probability measure on the set of distribution functions with countable support, with probabilities $\omega_j,j=1,2,\ldots$ at locations $x_j, j=1,2, \ldots \in \mathcal{X}$, and the $DP(\alpha, G)$ model induces randomness by drawing the $\omega_j$s via a probabilistic algorithm that depends on $\alpha$ -- commonly the so-called `stick-breaking' algorithm is used -- and the $x_j$ independently from $G$. In the most common form of Bayesian non-parametric analysis, the Dirichlet process acts as a prior for parameter $F_O$; hyperparameter $\alpha > 0$ acts as a concentration parameter, and $G(.)$ is a prior (base) distribution with domain $\mathcal{X}$.  In light of data $o_{1:n}$, the resulting posterior distribution is also a Dirichlet process $DP(\alpha_n,G_n)$ where $\alpha_n = \alpha+n$ and $G_n(\ldotp) =  w_n G(\ldotp) + (1-w_n) \widehat F_n(\ldotp)$, where $w_n = \alpha/(\alpha+n)$ and $\widehat F_n(\ldotp)$ is the empirical measure derived from $o_{1:n}$.

It is straightforward to generate samples from $DP(\alpha_n,G_n)$ (that is, randomly generated distributions that represent sampled versions of `parameter' $F_O$) and also from the implied model for the observable quantities in light of the data (that is, a randomly generated posterior predictive distribution).  Furthermore, the Dirichlet process posterior becomes concentrated at the data generating model $F_0$ in the limit as $n \longrightarrow \infty$ \citep[section 4.7]{ghosal_van_der_vaart_2017}, and provides a consistent estimation procedure.

With this relaxation of the parametric assumption about the data generating model, the calculations from section \ref{sec:Bims} can be reproduced.  The Bayes estimate again results from a minimum loss calculation based on the posterior predictive distribution.  When the posterior distribution is the $DP(\alpha_n,G_n)$ distribution, we have, for example replicates $\vartheta^{(l)}, l=1,\ldots,L$ sampled from the posterior for $\vartheta$ given by
\begin{equation}\label{eq:PostPredEstLogMisspecSampDP}
\vartheta^{(l)} = \arg \max_{t^\prime \in \Theta^\prime} \sum_{j=1}^\infty \omega_j^{(l)} \log f(s_j^{(l)};t^\prime)
\end{equation}
where $\{\omega_j^{(l)}, j=1,2,\ldots\}$ are a sample of probabilities drawn by, say, stick-breaking with parameter $\alpha_n$, and $\{s_j^{(l)}, j=1,2,\ldots\}$ are drawn independently from $G_n$.  In practice, the infinite sum is truncated by machine accuracy, as the $\omega_j$ values decrease in expectation as $j$ increases.  The $\{\omega_j\}$ may also be drawn such that they are decreasing in magnitude, rendering the truncation straightforward to implement.

\subsection{The Bayesian bootstrap}

The Bayesian bootstrap posits a multinomial likelihood on the finite set $\mathcal{O} = \{o_1,\ldots,o_n\}$ with unknown probabilities $\varpi = (\varpi_1,\ldots,\varpi_n)$ attached to each element, and combines this with a $Dirichlet(\alpha,\ldots,\alpha)$ prior to yield the posterior distribution for $\varpi$ to be $Dirichlet(\alpha+1,\ldots,\alpha+1)$.  Taking $\alpha \longrightarrow 0$ yields the Bayesian bootstrap, in which the predictive distribution is represented
\begin{equation}\label{eq:DPform}
p_n(o) = \sum_{i=1}^n \omega_i \delta_{o_i}(o)
\end{equation}
where $\omega \sim Dirichlet(1,\ldots,1)$, identical to the posterior distribution.

In \cite{rubin1981bayesian}, the Bayesian bootstrap is proposed as a heuristic strategy, but its theoretical properties have since been widely studied; see for example \cite{Lo1987,ChengHuang2010} and \cite{ghosal_van_der_vaart_2017}. The argument confirming that this strategy was in fact producing an approximate Bayesian posterior statements was formalized by \cite{newton1994approximate}.  The Newton \& Raftery algorithm is central to the procedures used in the Bayesian causal settings in \cite{saarela2015bayesian} and \cite{saarela2016}: in those papers, the utility argument is made explicit, and the log-density utility is justified by considering a hypothetical \textit{experimental} data generating mechanism that is explicitly misspecified (compared to the \textit{observational} data generating model).  See also \cite{chamberlainimbens2003} and \cite{GMS2016} for examples, and \cite{lyddon2019} for some generalizations.

The Bayesian bootstrap results as is the consequence of a Dirichlet process specification for the probability model that generated data $o_{1:n}$, in the limiting case $\alpha \longrightarrow 0$. Sampling from the posterior predictive coincides with the Bayesian bootstrap; if $\omega = (\omega_1,\ldots,\omega_n) \sim Dirichlet(1,1,\ldots,1)$, \eqref{eq:DPform} yields the estimation procedure
\begin{equation}\label{eq:BBsample}
\vartheta = \arg \max_{t^\prime \in \Theta^\prime} \sum_{i=1}^n \omega_i \log f(o_i;t^\prime)
\end{equation}
with $\vartheta$ now being a random quantity as $\omega$ is random. The summation in this expression is a \textit{deterministic} function of $\omega$ for every fixed $t^\prime$;  therefore the corresponding $\vartheta$ is also a deterministic function of $\omega$.  Hence, once we have sampled the weights in the Dirichlet process formulation, a transformation yields $\vartheta$, and thus $\vartheta$ is simply a functional of the Dirichlet process posterior on $F_O$.  Therefore the posterior sample formed by repeatedly sampling the Dirichlet weights to yield $\omega^{(1)},\ldots,\omega^{(L)}$, with subsequent transformations to yield $\vartheta^{(1)}, \ldots, \vartheta^{(L)}$ is an exact sample from the posterior distribution for $\vartheta$.  A proper prior $\pi_0(\vartheta)$ can be incorporated by modifying the specified utility function as in \eqref{eq:uPrior}.

Such inference is a fully Bayesian expression of posterior beliefs concerning the target of inference under the Bayesian non-parametric formulation.  As for any MCMC-based analysis, inference is only exact up to Monte Carlo sampling, that is, we can only compute the distribution of $\vartheta$ by sampling the Dirichlet process, and not analytically.  The calculation based on the formulation equivalent to \eqref{eq:EstEq} involves solving
\begin{equation}\label{eq:EstEqDP}
\sum_{j=1}^\infty \omega_j^{(l)} \dot{u}_\theta(s_j^{(l)},\vartheta) = 0
\end{equation}
where $(\omega_j^{(l)},s_j^{(l)}),j=1,2,\ldots$ define a random draw from the Dirichlet process posterior.

\subsection{Bayesian inference for the structured causal model}

For the causal inference problem with observed data $o_{1:n} = (x_{1:n},y_{1:n},z_{1:n})$, for a parametric analysis, we may compute the posterior distribution for $\theta = (\eta,\gamma,\zeta)$ using a factorization of the full model as in \eqref{eq:factorization01}.  We can also define the approximating model to respect the entire factorization, or target some component of interest.  For example, a conditional model for $Y$ given $(X,Z)$ might be targeted, with $u_\theta(o,\vartheta) = - \log f(y|x,z;\vartheta)$ for some conditional density $f(\ldotp;x,z;\vartheta)$.  Then, by sampling the posterior for $\theta$, or the posterior predictive distribution, the method of section \ref{sec:Bdti} can be deployed.

For the illustrative model of section \ref{sec:im}, let $\theta = (\xi,\tau)$ and $\vartheta = (\phi,\tau)$ be the parameters in the data generating and approximating models respectively. In this parametric setting,  assuming Normally distributed residual errors in both models, $\pi_n(\theta)$ is readily computable, and using the methods described in section \ref{sec:Bdti} we can obtain a sample from the posterior distribution and estimate for $\vartheta$ in the approximating model.   Specifically, from the model \eqref{eq:RMN1}, we have for $b(\ldotp)$ known $u_\theta(o,\vartheta) = ((y - b(x) \phi - z \tau)/\lambda)^2$.  In this case the parameter of interest $\tau$ is identical in the two models, and the posterior computed for $\pi_n(\theta)$ yields correct inference under the presumed correct specification of the conditional model.  The posterior for $\tau$ as a component of $\vartheta$ would still concentrate at true value $\tau_0$, but in finite sample the posterior variance would be larger than that computed from the correctly specified model that led to $\pi_n(\theta)$.

To relax the assumption of Normal residual errors in the data generating model, we may use the Bayesian bootstrap, and obtain a sampled variate from the posterior as
\begin{equation}\label{eq:RMN1-BB}
(\phi^{(l)},\tau^{(l)}) = \arg \min_{(\phi,\tau)} \sum_{i=1}^n \omega_j^{(l)} (y_i - b(x_i) \phi - z \tau)^2
\end{equation}
for which the minimization can be achieved analytically for $l=1,\ldots,L$.  

In \eqref{eq:RMN1-BB}, the Bayesian bootstrap is being used to sample the Dirichlet process posterior for the entire unknown joint distribution of the observables, but in the approximating parametric model only the conditional distribution for $Y$ given $X$ and $Z$ is studied -- the joint distribution does correspond to an implied conditional distribution.  This possibility of partial specification of the model of interest is an advantage of the formulation from section \ref{sec:Bdti}.  In addition, if the utility is modified to be
\[
u_\theta(o,\vartheta) = - \log f_1(y|x,z;\vartheta_1) - \log f_2(z|x;\vartheta_2)
\]
for proposed conditional densities $f_1$ and $f_2$.  Estimation or posterior sampling of $\vartheta_1$ and $\vartheta_2$ using the parametric or non-parametric algorithms can proceed by the obvious extension, and in this separable loss function the two optimizations can be carried out separately.  However, in the inference problem for \eqref{eq:RMN1} with propensity score unknown, a modification of the loss function is required for optimal inference. Suppose that
\begin{equation}\label{eq:uopt}
u_\theta(o,\vartheta) = - \log f_1(y|x,z;\vartheta_1,\vartheta_2^\opt) - \log f_2(z|x;\vartheta_2)
\end{equation}
where $\vartheta_2^\opt$ is the loss minimizing value of $\vartheta_2$ obtained by considering the second term only.  This utility reflects the estimation task in the causal problem based on \eqref{eq:RMN1}; the outcome model based on $f_1$ is adjusted using the fitted propensity score computed using the best estimate of the data generating parameter in the model $f_2$.

Taking \eqref{eq:RMN1} or \eqref{eq:RMN2} as the approximating model, inference for $\vartheta_1 = (\beta,\phi,\tau)$ will be correct (specifically consistent for, and with the posterior concentrated at, true value $\tau_0$) provided the propensity score model encapsulated in model $f_2$ is itself correctly specified with $\vartheta_2 \equiv \gamma$, so that the estimated propensity score based on the posterior mode $f_2(z|x;\widehat \vartheta_2)$ consistently estimates the true propensity score.

\section{Conventional Bayesian propensity score adjustment}\label{sec:FormalFrame}

Underlying our concept of a valid Bayesian approach is one which relies on the de Finetti representation for observable quantities in the data generating model as in section \ref{sec:Exchangeability}, with inference following a decision-theoretic argument as in section \ref{sec:Bims}.  It is common, however, to apply the Bayesian logic to procedures such as those indicated in section \ref{sec:Bm}.  Such procedures also can be assessed as fully Bayesian by reference to the decision-theoretic formulation of section \ref{sec:Bdti}.

\subsection{Joint estimation}
\label{sec:jointjust}

Estimation using the joint Bayesian model in \eqref{eq:jointLikelihood} can be justified using either conventional Bayesian logic or the arguments in section \ref{sec:Bdti} leading to the Gibbs posterior formulation and \eqref{eq:GibbsProblem}, that is, with
\begin{equation}\label{eq:bobjointRMN}
\ell \left(o_{1:n},(\beta,\phi,\tau,\gamma) \right) = - \sum_{i=1}^n \log f_{Z|X}(z_i|x_i;\gamma) - \sum_{i=1}^n \log f_{Y|X,Z}(y_i|x_i,z_i;\beta,\phi,\tau,\gamma).
\end{equation}
However, the resulting posterior does not concentrate at the correct ATE due to `feedback' which arises because the outcome depends on the parameters associated with the exposure model. A graphical model argument can be made to support this.  Feedback is present because of a `backdoor' path \citep{galles1995testing} from $\gamma$ to $(\beta,\phi,\tau)$ via $B$ in the graph describing the joint distribution of parameters and observables if the dependence of $Y$ on the confounders is mis-specified; $B$ is a `collider' on this path, so conditioning on it opens the path. As a result, the propensity score estimated in this way will not have the balancing property, even as $n$ increases.

A Bayesian analysis based on \eqref{eq:bobjointRMN} may, of course, still be carried out, and in finite sample the performance of the resulting Bayesian inference summaries may be acceptable; for example, the resulting estimators may have low variance.  However, as the sample size grows, it is clear from classical arguments that the Bayesian estimator of $\tau$ will be inconsistent.

\subsection{Cutting feedback and two-step estimation}
\label{sec:mea}

As noted in section \ref{sec:role}, $B$ should be constructed as $b(X;\gamma_0)$, and if $\gamma_0$ is unknown, it should be estimated using the observed $X$ and $Z$ values only.    The conventional Bayesian analysis therefore should be based on the posteriors
\begin{align}
\pi_n(\gamma) & \propto f_{Z|X}(z_{1:n}|x_{1:n};\gamma)\pi_0(\gamma) \label{eq:correct1}\\
\pi_n(\beta,\phi,\tau) & \propto f_{Y|X,Z}(y_{1:n}|x_{1:n},z_{1:n},b(x_{1:n};\gamma_0);\beta,\phi,\tau)\pi_0(\beta,\phi,\tau) \label{eq:correct2}
\end{align}
where $\gamma_0$ is the degenerate limiting value of $\pi_n(\gamma)$ referred to in section \ref{sec:Exchangeability}. We first compute the posterior for $\gamma$ from \eqref{eq:gammaposterior}, then we compute a Bayesian estimate $\widehat \gamma$ and fitted values $\widehat b_i = b({x}_i;\widehat \gamma)$ for $i=1,\ldots,n$.   The posterior distribution is computed via \eqref{eq:gammaposterior2S} and we can marginalize out to obtain $\pi_{n}(\tau)$..  Because of the conditioning on a specific $\gamma$ value, there is in fact no `feedback'.   The use of a plug-in estimate $\widehat \gamma$ may lead to imperfect adjustment for confounding in finite samples, but this is not due to feedback in the sense described above.

From the decision-theoretic perspective, a fully Bayesian justification via the variational formulation and \eqref{eq:GibbsProblem} is obtained using the loss function
\begin{equation}\label{eq:bobtwostageRMN}
\ell \left(o_{1:n},(\beta,\phi,\tau,\gamma) \right) = - \sum_{i=1}^n \log f_{Z|X}(z_i|x_i;\gamma) - \sum_{i=1}^n \log f_{Y|X,Z,B}(y_i|x_i,z_i, \widehat b_i ;\beta,\phi,\tau,\gamma^\opt)
\end{equation}
where $\gamma^\opt$ is itself a loss-minimizing quantity, say the posterior mode or mean, derived using the variational solution from \eqref{eq:bobj}, which under an independent prior specification is the conventional posterior for $\gamma$.  The formulation in \eqref{eq:bobtwostageRMN} evidently leads to a form of `modularized' inference as advocated by \cite{bayarri2009,zigler2016central} and \cite{jacob2017better}.    In this setting, however, due to the requirement to use a `best estimate' of $\gamma$ in order to produce consistent estimation of $\tau$, the modularization is a necessary step rather than a choice the Bayesian analyst may opt to make.

The cut feedback approach is an attempt to account for the uncertainty in estimating $\gamma$.  In this approach, $L$ samples $\gamma^{(1)},\ldots,\gamma^{(L)}$ from the posterior distribution $\pi_n(\gamma)$ are drawn, and each is used to compute a set of propensity score values, leading to $L$ parallel analyses that involved drawing a single sample from \eqref{eq:gammaposteriorCF}, the posterior computed using the $l$th sampled $\gamma$ value.  However, recall that only if $\gamma = \gamma_0$ do we achieve the required balance.  Thus when the propensity score values $b_{1:n}^{(l)}$ are computed using $\gamma^{(l)}$, they can be interpreted as error-corrupted versions of the true balancing scores $b_i = b(x_i;\gamma_0)$, and hence will not induce balance. Using a Taylor expansion, we have
\[
b_{i}^{(l)} \bumpeq b_i + \dot{b}(x_i;\gamma_0) (\gamma^{(l)} - \gamma_0) = b_i + u_i^{(l)}(x_i)
\]
say, where $\dot{b}(x;\gamma)$ is the partial derivative of $b(x;\gamma)$ taken with respect to $\gamma$. Hence when $b_{i}^{(l)}$ is used in the propensity score regression approach, we should regard it as an error-corrupted version of the balance-inducing (but unknown) value $b_i$, where the error has variance proportional to the (posterior) variance of the sampled values $\gamma^{(l)}$.  It is well-known that the presence of such error in regressors in a regression model typically leads to bias in the estimation of regression coefficients even if the functional form of the model is correct -- although unlike the commonly-cited setting that leads to attenuation, here the measurement errors $u_i^{(l)}(x_i)$ are dependent on the observed $x_i$, and thus have different variances.  A numerical example to illustrate the bias induced by the cut-feedback procedure is given in Appendix \ref{sec:AppBias}.


\subsection{Frequentist assessment of the conventional Bayesian estimators}\label{sec:Xvar}

The presence of $\gamma_0$ in \eqref{eq:correct2} in practice requires the use of a Bayesian estimate to facilitate computation.  A natural estimator is the posterior mean or mode derived from \eqref{eq:correct1}, and plugging the corresponding estimate into \eqref{eq:correct2} allows posterior inference to proceed.  Even if no account is taken of the estimation of $\gamma$, then the analysis of the parameters in the outcome model is being performed in a standard Bayesian fashion.  It is evident from \eqref{eq:correct1} and \eqref{eq:correct2} that $\gamma$ and $(\beta,\phi,\tau)$ are \textit{a posteriori} independent.  Therefore plugging in an estimate $\widehat \gamma$  -- a deterministic function of the $(x,z)$ data -- derived from $\pi_n(\gamma)$ has no impact on inference for $(\beta,\phi,\eta)$ provided the treatment model is correctly specified.

There are two things to note about this procedure.  First, in finite sample, the posterior variance for $\tau$ is smaller when using an estimate of $\gamma$ rather than the true value if it were known, in a result that is analogous to the results in the frequentist literature from \cite{HIR2003} and \cite{henmieguchi2004}; see the results in Appendix Table \ref{tab:sim1tab} and related discussion. Secondly, if the plug-in approach is adopted, the resulting Bayesian inference exhibits relatively poor frequency properties: across replicate data sets of the same size, coverage properties of Bayesian credible intervals derived from $\pi_n(\beta,\phi,\eta)$ with $\gamma_0$ estimated by $\widehat \gamma$ are below the nominal level.  This phenomenon arises from the fact that the model for the data generating process is mis-specified, and therefore frequentist behaviour (across replicate data sets) is not adequate. If inference is made using the posterior distribution conditioned on the observed data, standard Bayesian inference methods under exchangeability and correct specification (that is, that follow the de Finetti representation) will have expected frequentist bahaviour.  However, if the presumed data generating process is mis-specified, then we have no such guarantees.  This issue is overcome by the use of the Bayesian non-parametric model and the Bayesian bootstrap.  See the simulation study in Appendix \ref{sec:undercover}.

\section{Simulation studies}\label{SimStudy}

We examine the performance of the conventional Bayesian computational methods described in section \ref{sec:Bm} with the decision-theoretic and non-parametric methods from sections \ref{sec:Bdti} and \ref{sec:BBootsec}.
		
\subsection{Example 1:  Normal exposure}
\label{sec:Ex1}

In this simulation, the data generating mechanism assumes $p=3$ confounders, with $x = (x_1,x_2,x_3)^\top \sim Normal((-1,2,0.5)^\top, \Sigma)$, with $\Sigma_{ij} = \textrm{Cov}(X_i,X_j) = 0.8^{|i-j|}$, for $i,j = 1,2,3$. We consider sample sizes $n = 200, 500, 1000$ and $2000$, and simulate $Z_i$ and $Y_i$ from Normal distributions with unit variance and means
\begin{align*}
\mu_{Z} &= 1  - x_{1} + x_{2} + 2x_{3} - x_{1}x_{2} + 2x_{2}x_{3} ,
\\[3pt]
\mu_{Y} &= 1  + 5 z  + x_{1} + x_{2} + x_{3} + 5x_{2}x_{3}.
\end{align*}
respectively.  For each sample size, we generate $1000$ datasets under the above scheme.  For the exposure model, we fit the mean model $\mu_{Z} = \widetilde x {\gamma}$, where the linear predictor is based on $\widetilde x = (1,x_1,x_2,x_3,x_1x_2,x_1x_3, x_2x_3,x_1x_2x_3)^\top$, using linear regression.

\subsubsection{Conventional Bayesian methods}
	
We fitted several parametric models under the assumption of Normal errors.  In the cutting feedback models, $\tilde{b}_i = \widetilde x_i \tilde{\gamma}$ with $\tilde{\gamma}$ being the sampled value of $\gamma$ in a Gibbs sampler procedure, and in the two-step models $\widehat{b}_i = \widetilde x_i \widehat{\gamma}$, where $\widehat{\gamma}$ is the Bayesian estimator of $\gamma$ obtained from the fitted exposure model.
\vspace{-2.5mm}
\begin{itemize}
\setlength\itemsep{-1.0em}
  \item `Unadjusted (UN)': unadjusted for confounding;
  \begin{eqnarray*}
\\[-18pt]
				\textrm{UN}:& & \beta_0 + \tau z \\[3pt]
				\textrm{UN-ext}: & & \beta_0 + x_1 \beta_1 + x_2 \beta_2 + x_3 \beta_3 + \tau z
\end{eqnarray*}

  \item `Joint (JT)': the joint model from equation \eqref{eq:jointLikelihood};
  \begin{eqnarray*}
\\[-18pt]
				\textrm{JT}: & & \beta_0 + \phi \widetilde x {\gamma} + \tau z\\[3pt]
				\textrm{JT-ext} : & & \beta_{0} + x_1 \beta_1 + x_2 \beta_2 + x_3 \beta_3 + \phi \widetilde x {\gamma} + \tau z
\end{eqnarray*}

  \item `Cutting feedback (CF)': the cut feedback approach via equation \eqref{eq:gammaposterior}
  \begin{eqnarray*}
\\[-18pt]
				\textrm{CF} : &  & \beta_{0} + \phi \tilde{b} + \tau z \\[3pt]
				\textrm{CF-ext} : &  & \beta_{0} + x_1 \beta_1 + x_2 \beta_2 + x_3 \beta_3 + \tau z + \phi \tilde{b}
\end{eqnarray*}

  \item `Two-step (2S)':
  \begin{eqnarray*}
\\[-18pt]
				\textrm{2S} : &  &  \beta_{0} + \phi \widehat{b}+ \tau z \\[3pt]
				\textrm{2S-ext} : & & \beta_0 + x_1 \beta_1 + x_2 \beta_2 + x_3 \beta_3 + \phi \widehat{b} + \tau z
\end{eqnarray*}

  \item `Correct': a correct specification of the linear regression model.
\end{itemize}
\vspace{-2.5mm}
	
	\begin{table*}[ht]
		\centering
		\caption{Simulated Example 1: Summary of the conventional Bayesian estimates of $\tau$ under a normal exposure. The rows correspond to mean bias of the point estimates, RMSE and the coverage rates of the posterior 95\% credible intervals of $\tau$. Results over 1000 replicate data sets.}
		\label{tab:normalExpStudy}
		\begin{tabular}{llrrrr}
\\[-3pt]
	\hline
    &         & \multicolumn{4}{c}{$n$} \\
	& Outcome & \multicolumn{1}{c}{$200$}& \multicolumn{1}{c}{$500$} & \multicolumn{1}{c}{$1000$} & \multicolumn{1}{c}{$2000$}\\
\hline
\multirow{8}{*}{\rotatebox[origin=c]{90}{Bias}} &UN & 2.084 & 2.092 & 2.093&  2.089\\
&UN-ext &  2.401&  2.448&  2.444&  2.444\\
&JT     & -0.355& -0.345& -0.344& -0.345\\
&JT-ext & -0.092& -0.088& -0.089& -0.090\\
&CF     &  0.059&  0.027&  0.013&  0.006\\
&CF-ext &  0.045&  0.021&  0.011&  0.005\\
&2S     & -0.002&  0.001&  0.001&  0.000\\
&2S-ext & -0.002&  0.001&  0.001&  0.000\\
&Correct& -0.002&  0.001& -0.001&  0.000\\
				\hline
\multirow{8}{*}{\rotatebox[origin=c]{90}{RMSE}}&UN & 2.086&0.093&2.093&2.089\\
&UN-ext & 2.416& 2.454& 2.447& 2.445\\
&JT     & 0.365& 0.349& 0.346& 0.346\\
&JT-ext & 0.117& 0.100& 0.095& 0.093\\
&CF     & 0.092& 0.054& 0.035& 0.024\\
&CF-ext & 0.084& 0.051& 0.034& 0.023\\
&2S     & 0.071& 0.047& 0.033& 0.023\\
&2S-ext & 0.071& 0.047& 0.033& 0.023\\
&Correct& 0.056& 0.036& 0.025& 0.018\\
				\hline
\multirow{8}{*}{\rotatebox[origin=c]{90}{Coverage}}&UN & 0.0 & 0.0 & 0.0 & 0.0\\
&UN-ext &   0.0 &   0.0 &   0.0 &  0.0\\
&JT     &   0.1 &   0.0 &   0.0 &  0.0\\
&JT-ext &  75.0 &  49.7 &  19.8 &  2.1\\
&CF     & 100.0 & 100.0 & 100.0 & 100.0\\
&CF-ext & 100.0 & 100.0 & 100.0 & 100.0\\
&2S     & 100.0 & 100.0 & 100.0 & 100.0\\
&2S-ext & 100.0 & 100.0 & 100.0 & 100.0\\
&Correct&  94.1 &  94.5 &  94.1 &  94.0\\
				\hline
		\end{tabular}
	\end{table*}

Table \ref{tab:normalExpStudy} contains the estimated bias and root mean square error (RMSE) the posterior estimates (means), and coverage of the $95\%$ credible interval) for $\tau$.  The unadjusted and joint models perform poorly as theory suggests.  Estimation based on cutting feedback yields a small amount of bias, which decreases as the sample size $n$ increases. The two-step approaches yield unbiased estimators.  However, in all cases the coverage of the Bayesian credible intervals is not adequate when the outcome model is mis-specified, even though coverage at the nominal level can be obtained using a correct specification.

\subsubsection{Estimation via the Bayesian bootstrap}

The results demonstrate that model mis-specification disrupts parametric Bayesian inference.  We repeated the analysis using the Bayesian bootstrap approach, restricting attention to the cutting feedback and two-step estimation procedures.  To implement the cutting feedback procedure, recall that the Bayesian bootstrap produces a sample from the posterior for a target parameter.  In our analysis, we assume correct specification for the treatment assignment model, and so for the posterior for $\pi_n(\gamma)$, we may either use the exact posterior computed under a Normal assumption, or the Bayesian bootstrap.  Having obtained a sample of size $L$ from this posterior, we then use the Bayesian bootstrap to generate $L$ posterior samples for $\tau$, conditioning on the fitted value $\tilde{b}_i = \widetilde x_i \tilde{\gamma}$.  For the two-step method, we may proceed in the same fashion, but instead use $\widehat{b}_i = \widetilde x_i \widehat{\gamma}$, where $\widehat{\gamma}$ is the posterior mean derived from $\pi_n(\gamma)$.

These methods follow the conventional approach of separating the posteriors from the two parts of the model.  However, following the argument leading to \eqref{eq:uopt}, the correct Bayesian approach retains the linkage of the two models via the common Dirichlet weights noted in \eqref{eq:BBsample}; that is, a \textbf{single} draw of weights $\omega$ is used in the optimization over $\gamma$ and the consequent optimization over $(\beta,\phi,\tau)$.   This linkage reflects a Bayesian non-parametric assumption concerning the full joint distribution of the observables.

	\begin{table*}[ht]
		\centering
		\caption{Simulated Example 1: Summary of the estimates of $\tau$ under a normal exposure using the Bayesian bootstrap in the outcome model, and different approaches to the propensity score model parameters posterior: True indicates the true value of $\gamma$ is used; Parametric indicates a parametric Normal model is used;  Unlinked indicates that the posteriors for $\gamma$ and $(\beta,\phi,\tau)$ were computed using separate Bayesian bootstrap computations and different Dirichlet weights (Unlinked Bayesian bootstrap, UBB); Linked (LBB) indicates that common Dirichlet weights were used in the two model components. Rows correspond to RMSE and the coverage rates of the posterior 95\% credible intervals. Results over 1000 replicate data sets.}
		\label{tab:BootNorm}
		\begin{tabular}{lllrrrr}
\\[-3pt]
	\hline
    &         &          & \multicolumn{4}{c}{$n$} \\
	& Outcome & $\pi_n(\gamma)$ & \multicolumn{1}{c}{$200$}& \multicolumn{1}{c}{$500$} & \multicolumn{1}{c}{$1000$} & \multicolumn{1}{c}{$2000$}\\
\hline
\multirow{12}{*}{\rotatebox[origin=c]{90}{RMSE}}
&PS          & True       &0.417&0.272&0.194&0.132\\
&PS-ext      & True       &0.214&0.143&0.096&0.069\\
&CF          & Parametric &0.093&0.056&0.035&0.024\\
&CF-ext      & Parametric &0.084&0.052&0.035&0.023\\
&2S          & Parametric &0.073&0.048&0.032&0.023\\
&2S-ext      & Parametric &0.072&0.047&0.032&0.022\\
&CF          & Unlinked BB&5.487&3.518&2.532&1.757\\
&CF-ext      & Unlinked BB&0.083&0.052&0.034&0.023\\
&2S          & Unlinked BB&0.078&0.050&0.033&0.022\\
&2S-ext      & Unlinked BB&0.072&0.048&0.032&0.022\\
&2S          & Linked BB  &0.071&0.047&0.032&0.022\\
&2S-ext      & Linked BB  &0.071&0.047&0.032&0.022\\
\hline
\multirow{12}{*}{\rotatebox[origin=c]{90}{Coverage}}
&PS          & True       & 94.2&  94.0& 95.0&  96.0\\
&PS-ext      & True       & 93.1&  92.8& 94.1&  94.8\\
&CF          & Parametric &100.0& 100.0&100.0& 100.0\\
&CF-ext      & Parametric &100.0& 100.0&100.0& 100.0\\
&2S          & Parametric &100.0& 100.0&100.0& 100.0\\
&2S-ext      & Parametric &100.0& 100.0&100.0& 100.0\\
&CF          & Unlinked BB& 96.5&  95.3& 94.1&  95.1\\
&CF-ext      & Unlinked BB&100.0& 100.0&100.0& 100.0\\
&2S          & Unlinked BB&100.0& 100.0&100.0& 100.0\\
&2S-ext      & Unlinked BB&100.0& 100.0&100.0& 100.0\\
&2S          & Linked BB  & 94.2&  92.8& 94.7&  94.1\\
&2S-ext      & Linked BB  & 94.2&  92.8& 94.7&  94.1\\
\hline
		\end{tabular}
	\end{table*}

For the treatment assignment model, we carry out analysis using the \textit{\textbf{True}} propensity score, and then compute $\pi_n(\gamma)$ using a \textbf{\emph{Parametric}} (logistic regression) analysis, using the Bayesian bootstrap in an \textbf{\textit{Unlinked}} fashion (via independent Dirichlet weights in the two components of expression \eqref{eq:uopt}), and in a \textbf{\textit{Linked}} fashion using a single Dirichlet draw.  For the outcome model, we use a least-squares optimization for the Bayesian bootstrap sampling of $(\beta,\phi,\tau)$.  The analyses were conducted in 1000 replicate data sets, using $1000$ Bayesian bootstrap draws for each replicate.   For each data replicate, we compute the RMSE of the Bayesian posterior estimates;  coverage rates were computed by constructing, for each replicate data set, posterior sample quantiles. The results are presented in Table \ref{tab:BootNorm}.  All of the methods were unbiased in large sample, although the CF method showed a small bias as discussed in section \ref{sec:mea} when $n$ was small, and also larger variability.  In terms of RMSE, the two-step methods generally performed best.  Coverage at the nominal level was recovered for the two-step method in a Linked analysis, as suggested by the theory studied in section \ref{sec:BBootsec}.

	\subsection{Example 2: Binary exposure}
\label{sec:Ex2}
	
In example 2, we consider $p = 4$ independent confounders, $X=(X_1,X_2,X_3,X_4)$, with $X_1,X_2 \sim Normal(1,1)$, and $X_3, X_4 \sim Normal(-1,1)$, and specify different propensity score distributions, to investigate how this distribution affects the estimation of the treatment effect.  For each $i$, $Z_i \sim Bernoulli(p_i)$ with
\begin{equation}
\mbox{logit} (p_i) = \gamma_0 + \gamma_1 x_{i1} + \gamma_2 x_{i2} + \gamma_3x_{i3} + \gamma_4x_{i4},
\label{expSimStudy}
\end{equation}
and three settings of the parameters: Scenario 1 $\gamma = (0.0,0.3,0.8,0.3,0.8)$; Scenario 2 $\gamma = (0.5,0.5,0.75,1.0,1.0)$; Scenario 3 $\gamma = (0.0,0.45,0.90,1.35,1.8)$.  Scenario 1 has a reasonably uniform distribution of propensity scores, Scenario 2 has a slight preponderance of lower scores, and Scenario 3 has very few high scores.  For the outcome model, we simulated $Y \sim Normal(0.25x_{1} + 0.25x_{2} + 0.25x_{3} + 0.25x_{4} + 1.5x_{3}x_{4},1)$.  Under this scenario there is no treatment effect. In the analyses, the exposure model is correctly specified via (\ref{expSimStudy}), and for the outcome model, we fit the same models as in section \ref{sec:Ex1}.  We assign $Normal(0,10^2)$ priors for the elements of $\gamma$, but non-informative priors for the parameters in the outcome model.

The results in Table \ref{tab:Ex2RMSE} suggest that the distribution of the propensity score affects the performance of all methods.   For Scenarios 1 and 2, for small values of $n$, two-step methods again perform better than CF methods in terms of RMSE due to the finite sample bias.  Scenario 3 is the case where the true propensity score distribution is most skewed; overall, CF methods show similar results to 2S in terms of RMSE for $n=100$, and $200$.  In terms of coverage, for all scenarios, CF and 2S models show coverage rates that are always above the target level, even though the target coverage is achievable under correct specification.
	
\begin{table}[ht]
\setlength{\tabcolsep}{0.25em}
\centering
	\caption{Simulated example 2: RMSE of the estimates of $\tau$, and coverage of the 95\% credible interval for a binary exposure model.  Rows correspond to the averages of the posterior means and variances, and coverage rates of $\tau$ for three settings of the propensity score model given in section \ref{sec:Ex2}. Results from 1000 replicate data sets.\label{tab:Ex2RMSE}}
\smallskip
\begin{tabular}{l|rrrr|rrrr|rrrr|}
\multicolumn{1}{l}{RMSE}& \multicolumn{4}{c|}{Scenario 1} & \multicolumn{4}{c|}{Scenario 2} & \multicolumn{4}{c|}{Scenario 3}\\
  \hline
 $n$ & 200 & 500 & 1000 & 2000 & 200 & 500 & 1000 & 2000 & 200 & 500 & 1000 & 2000 \\
  \hline
UN & 0.923 & 0.883 & 0.872 & 0.857 & 1.463 & 1.420 & 1.423 & 1.419 & 1.603 & 1.604 & 1.587 & 1.584 \\
  UN-ext & 0.263 & 0.174 & 0.125 & 0.089 & 0.306 & 0.197 & 0.142 & 0.116 & 0.746 & 0.685 & 0.667 & 0.653 \\
  JT & 0.512 & 0.417 & 0.382 & 0.366 & 0.615 & 0.529 & 0.485 & 0.477 & 0.938 & 0.819 & 0.784 & 0.763 \\
  JT-ext & 0.263 & 0.176 & 0.124 & 0.092 & 0.309 & 0.201 & 0.175 & 0.151 & 0.394 & 0.277 & 0.227 & 0.202 \\
  CF & 0.278 & 0.178 & 0.124 & 0.088 & 0.310 & 0.174 & 0.127 & 0.088 & 0.333 & 0.205 & 0.149 & 0.107 \\
  CF-ext & 0.259 & 0.172 & 0.122 & 0.088 & 0.278 & 0.171 & 0.122 & 0.087 & 0.361 & 0.210 & 0.151 & 0.105 \\
  2S & 0.263 & 0.173 & 0.122 & 0.088 & 0.281 & 0.169 & 0.123 & 0.088 & 0.337 & 0.203 & 0.149 & 0.107 \\
  2S-ext & 0.263 & 0.172 & 0.122 & 0.088 & 0.277 & 0.168 & 0.122 & 0.087 & 0.323 & 0.200 & 0.145 & 0.103 \\
  Correct & 0.159 & 0.102 & 0.075 & 0.052 & 0.181 & 0.105 & 0.074 & 0.055 & 0.209 & 0.130 & 0.092 & 0.063 \\
   \hline
\multicolumn{1}{l}{} & \\
\multicolumn{1}{l}{Coverage} & \\
\hline
UN & 32.7 & 2.3 & 0.0 & 0.0 & 1.1 & 0.0 & 0.0 & 0.0 & 0.6 & 0.0 & 0.0 & 0.0 \\
  UN-ext & 97.2 & 95.7 & 95.3 & 96.1 & 95.3 & 94.8 & 93.9 & 90.4 & 56.5 & 17.4 & 1.6 & 0.0 \\
  JT & 76.0 & 57.3 & 27.2 & 4.6 & 61.5 & 31.3 & 9.0 & 0.1 & 36.4 & 11.1 & 0.7 & 0.0 \\
  JT-ext & 94.8 & 94.3 & 95.7 & 73.4 & 36.6 & 22.7 & 18.2 & 24.9 & 64.3 & 68.7 & 67.3 & 50.6 \\
  CF & 100.0 & 99.8 & 99.9 & 99.9 & 99.3 & 100.0 & 99.8 & 99.8 & 99.7 & 100.0 & 99.8 & 99.9 \\
  CF-ext & 97.9 & 96.0 & 95.8 & 95.5 & 97.4 & 97.9 & 97.7 & 97.8 & 97.2 & 97.9 & 97.4 & 96.9 \\
  2S & 99.7 & 99.4 & 99.1 & 99.3 & 98.9 & 99.3 & 99.6 & 99.6 & 100.0 & 99.9 & 99.6 & 99.9 \\
  2S-ext & 97.5 & 95.5 & 96.2 & 96.1 & 96.7 & 97.9 & 98.0 & 97.5 & 97.9 & 97.7 & 97.2 & 97.1 \\
  Correct & 94.4 & 94.5 & 94.0 & 94.5 & 93.1 & 96.2 & 95.8 & 94.2 & 95.4 & 94.6 & 94.6 & 94.5 \\
   \hline
\end{tabular}
\end{table}

\subsubsection{Estimation via the Bayesian bootstrap}
	
For the Bayesian bootstrap procedure, we use the same simulation design and $L=2000$ Bayesian bootstrap draws for each data set.  Table \ref{tab:binaryBoot} shows the RMSE and coverage rates. Results largely agree with those observed in Example 1 where the Bayesian bootstrap is used. Overall the CF and 2S approaches show similar values of RMSE and coverage, and for the coverage in particular the general performance of the bootstrap methods seems an improvement over the results for the conventional analyses from Table \ref{tab:Ex2RMSE}.

\begin{table}[ht]
\setlength{\tabcolsep}{0.25em}
\centering
		\caption{Simulated Example 2: Summary of the estimates of $\tau=0$ under a binary exposure over 1000 replicate data sets derived using the Bayesian bootstrap in the outcome model, and different approaches to computing the propensity score parameters posterior: True indicates the true value of $\gamma$ is used; Par. indicates a parametric Normal model is used;  UBB indicates that the posteriors for $\gamma$ and $(\beta,\phi,\tau)$ were computed using separate Bayesian bootstrap computations and different Dirichlet weights; LBB indicates that common Dirichlet weights were used in the two model components. Rows correspond to RMSE and the coverage rates of the posterior 95\% credible intervals.}
		\label{tab:binaryBoot}
\smallskip
\begin{tabular}{ll|rrrr|rrrr|rrrr|}
\multicolumn{2}{l}{RMSE}& \multicolumn{4}{c|}{Scenario 1} & \multicolumn{4}{c|}{Scenario 2} & \multicolumn{4}{c|}{Scenario 3}\\
\hline
 &$n$& 200 & 500 & 1000 & 2000 & 200 & 500 & 1000 & 2000 & 200 & 500 & 1000 & 2000 \\
 \hline
   PS           & True        & 0.361 & 0.232 & 0.163 & 0.119 & 0.359 & 0.234 & 0.172 & 0.128 & 0.765 & 0.687 & 0.677 & 0.661 \\
   PS-ext       & True & 0.275 & 0.171 & 0.120 & 0.088 & 0.281 & 0.176 & 0.128 & 0.089 & 0.335 & 0.211 & 0.152 & 0.107 \\
   CF           & Par.  & 0.278 & 0.171 & 0.119 & 0.087 & 0.294 & 0.177 & 0.127 & 0.088 & 0.334 & 0.209 & 0.153 & 0.106 \\
   CF-ext       & Par.  & 0.269 & 0.168 & 0.118 & 0.086 & 0.273 & 0.172 & 0.125 & 0.087 & 0.351 & 0.203 & 0.147 & 0.101 \\
   2S           & Par.  & 0.275 & 0.172 & 0.119 & 0.087 & 0.278 & 0.177 & 0.127 & 0.088 & 0.340 & 0.209 & 0.153 & 0.106 \\
   2S-ext       & Par.  & 0.271 & 0.168 & 0.118 & 0.086 & 0.274 & 0.172 & 0.125 & 0.087 & 0.321 & 0.201 & 0.146 & 0.101 \\
   CF           & UBB & 0.285 & 0.173 & 0.120 & 0.087 & 0.303 & 0.183 & 0.131 & 0.090 & 0.345 & 0.214 & 0.153 & 0.106 \\
   CF-ext       & UBB & 0.268 & 0.168 & 0.117 & 0.086 & 0.273 & 0.173 & 0.125 & 0.087 & 0.340 & 0.209 & 0.152 & 0.103 \\
   2S           & UBB & 0.276 & 0.170 & 0.119 & 0.087 & 0.281 & 0.176 & 0.127 & 0.088 & 0.344 & 0.211 & 0.153 & 0.106 \\
   2S-ext       & UBB & 0.271 & 0.168 & 0.118 & 0.086 & 0.275 & 0.172 & 0.125 & 0.087 & 0.324 & 0.202 & 0.146 & 0.101 \\
   2S           & LBB & 0.270 & 0.168 & 0.118 & 0.086 & 0.273 & 0.173 & 0.126 & 0.088 & 0.332 & 0.206 & 0.151 & 0.105 \\
   2S-ext       & LBB   & 0.269 & 0.167 & 0.117 & 0.086 & 0.271 & 0.171 & 0.125 & 0.087 & 0.316 & 0.198 & 0.145 & 0.101 \\
\hline
\multicolumn{2}{l}{} & \\
\multicolumn{2}{l}{Coverage} & \\
\hline
   PS           & True        & 94.0 & 94.6 & 95.3 & 94.5 & 94.0 & 95.7 & 94.5 & 94.8 & 93.9 & 94.9 & 94.5 & 94.8 \\
   PS-ext       & True & 93.1 & 94.1 & 94.9 & 94.0 & 93.0 & 94.4 & 94.5 & 94.9 & 93.5 & 95.0 & 93.7 & 94.8 \\
   CF           & Par.  & 100.0 & 99.4 & 99.3 & 99.1 & 98.7 & 98.9 & 98.8 & 98.9 & 98.7 & 98.8 & 98.0 & 98.2 \\
   CF-ext       & Par.  & 94.3 & 94.8 & 94.8 & 94.5 & 95.0 & 95.1 & 94.5 & 95.5 & 95.0 & 96.0 & 95.7 & 96.2 \\
   2S           & Par.  & 99.6 & 99.0 & 99.3 & 99.1 & 98.3 & 98.5 & 98.8 & 98.9 & 96.9 & 98.0 & 97.5 & 98.2 \\
   2S-ext       & Par.  & 93.7 & 94.7 & 94.8 & 94.5 & 94.2 & 94.7 & 94.5 & 95.5 & 95.3 & 95.5 & 95.0 & 96.2 \\
   CF           & UBB & 99.9 & 99.7 & 99.8 & 99.8 & 98.8 & 99.3 & 99.5 & 99.6 & 98.4 & 99.0 & 98.9 & 98.9 \\
   CF-ext       & UBB & 94.3 & 95.0 & 94.7 & 94.7 & 95.1 & 95.3 & 95.3 & 96.0 & 95.2 & 95.8 & 95.8 & 97.0 \\
   2S           & UBB & 99.6 & 99.2 & 99.3 & 99.1 & 98.2 & 98.6 & 98.7 & 98.7 & 96.9 & 97.7 & 97.8 & 98.0 \\
   2S-ext       & UBB & 93.6 & 94.6 & 94.8 & 94.5 & 94.2 & 94.7 & 94.5 & 95.5 & 95.0 & 95.4 & 95.2 & 96.2 \\
   2S           & LBB & 92.7 & 93.8 & 94.4 & 94.0 & 92.8 & 93.7 & 94.0 & 95.1 & 91.3 & 93.0 & 93.8 & 94.5 \\
   2S-ext       & LBB   & 92.7 & 93.9 & 94.5 & 93.8 & 92.7 & 94.0 & 94.0 & 94.8 & 92.1 & 93.5 & 93.7 & 95.0 \\
   \hline
\end{tabular}
\end{table}

\subsection{Example 3: Comparison with Bayesian Causal Forests} 

In this section, we compare results from the Bayesian approaches described in this paper with results obtained from the Bayesian Causal Forests (BCF) method \cite{hahn2020bayesian}.  The BCF approach is an example of flexible modelling based on Bayesian additive regression trees fitted using MCMC to infer potentially heterogeneous treatment effects.  The BCF model for binary treatment is based on the linear predictor
\begin{equation}\label{eq:bcfmodel}
\mu_i = \mu(x_i, b(x_i)) + \tau(x_i,b(x_i)) z_i
\end{equation}
with assumed homoscedastic Normal errors, and with functions $\mu(\ldotp,\ldotp)$ and $\tau(\ldotp,\ldotp)$ estimated via flexible Bayesian modelling.  The propensity score $b(x)$ in \eqref{eq:bcfmodel} is typically estimated as part of a separate Bayesian model.  The approach is implemented efficiently in the \texttt{R} package `\textbf{bcf}'.  The BCF method allows for more flexibility than models such as \eqref{eq:RMN1} or \eqref{eq:RMN2} that are typically used; recall that standard implementations require correct specification of the treatment effect model $\tau(x_i,b(x_i))$. We would therefore anticipate better performance of the standard implementations if the correct specification assumption holds.  Nevertheless, a comparison is potentially enlightening.

With predictors $X_1,X_2 \sim Normal(1,1)$, $X_3,X_4 \sim  Normal(-1,1)$, we simulate $Z \sim Bernoulli(p)$, with ${\rm logit}(p) = 0.45x_1 + 0.9x_2 + 1.35x_3 + 1.8x_4$, to simulate a binary treatment, and treatment-free outcome model
\[
\mu(x) = x_{1} + x_{2} + x_{3} + x_{4} + .75x_{1}x_{3} + .75x_{2}x_{4} + .75x_{1}x_{4} + .75x_{3}x_{4}.
\]
For the treatment effect model, we assume that in the data generating model version of \eqref{eq:bcfmodel} we have $\tau(x) = \psi_0 + \psi_1 x_1$ with $(\psi_0,\psi_1)=(1,2)$ which yields an average treatment effect of $\psi_0+\psi_1 \E[X_1] = 3$.  We compare the BCF approach with two-step approach with correctly specified treatment-effect model, that is, with mean
\[
\beta_0 + (\psi_0 + \psi_1 x_1) z + (\phi_0 + \phi_1 x_1) \widehat b
\]
fitted using the Linked Bayesian bootstrap.  For the fitted BCF model we assume the more general structure $Y = \mu(x,\widehat{b}(x)) + \tau(x_1,\widehat{b}(x)) z +  \epsilon$, where $\epsilon \sim Normal(0,\sigma^2)$.  In both analyses, the propensity score model is estimated under correct specification.  The \textbf{bcf} package outputs individual-level posterior contrasts which can be converted into population average quantities via the sample average
\[
\widehat \mu = \frac{1}{n} \sum_{i=1}^n \widehat \tau(x_i,\widehat{b}(x_i))
\]
which is computed for each posterior sample.

The results of this analysis are presented in Table \ref{tab:Ex3bcf}.  The BCF method displays a small amount of bias for smaller sample sizes, but typically has a smaller variance, and therefore ultimately a lower RMSE.  The two-step method using the Linked Bayesian bootstrap gives coverage at the target level, but the coverage of the BCF method is below the target level.  Again it should be stressed that the comparison is not entirely fair, as the BCF method does not assume a known functional form for the treatment effect model, and therefore is robust to mis-specification of the treatment effect model.  It is surprising that the variance of the BCF estimator is lower than that derived from the two-step method, but this phenomenon appears to persist in other settings (see Appendix section \ref{sec:AppBCF}).  We note, however, that the BCF approach, or any flexible outcome regression model, can also be included within a Bayesian bootstrap, and that because of the properties of the non-parametric procedure, good frequency properties can be recovered.  On average, the BCF method required three times the computational expenditure of a non-parallelized version of the Bayesian bootstrap approach.

\begin{table}[ht]
\centering
	\caption{Simulated example 3: Comparison of results for two-step fitted using the Linked Bayesian bootstrap, and the BCF method. Summary of 2000 Bayesian estimates and credible intervals.  Rows correspond to the bias, RMSE, and coverage rates. \label{tab:Ex3bcf}}
\begin{tabular}{crrrrr}
  \hline
  && \multicolumn{4}{c}{$n$}\\
& & 200 & 500 & 1000 & 2000 \\
  \hline
\multirow{2}{*}{Bias}     & 2S  & -0.013 & -0.005 & -0.005 & 0.005 \\
                          & BCF &  0.120 &  0.067 &  0.042 & 0.030 \\
  \hline
\multirow{2}{*}{RMSE}     & 2S  &  0.311 &  0.191 &  0.137 & 0.100 \\
                          & BCF &  0.320 &  0.182 &  0.119 & 0.088 \\
  \hline
\multirow{2}{*}{Coverage} &  2S  & 94.4   & 94.1   & 94.5   & 93.5 \\
                          &  BCF & 91.5   & 90.1   & 90.5   & 87.9 \\
   \hline
\end{tabular}
\end{table}

\section{Beyond regression adjustment in the Normal model}\label{sec:Extensions}

In this section, we identify a number of extensions to the causal adjustment approach based on regression, including situations where flexible modelling of the expected outcome conditional on treatment and confounders cannot recover the causal effect.

\subsection{Inverse probability weighting}
\label{sec:ipw}

Inverse probability weighting (IPW) is an alternate procedure for making causal adjustment based on the propensity score.  Inverse weighting breaks the confounding by converting the original sample into a pseudo-sample in which confounder imbalance is removed.  The loss/utility specification for this adjustment methods takes the form
\[
u_\theta(o,\vartheta) = - w(x,z;\vartheta_2^\opt) \log f_1(y|x,z;\vartheta_1) - \log f_2(z|x;\vartheta_2)
\]
where $w(x,z;\vartheta_2^\opt) = 1/f_2(z|x;\vartheta_2^\opt)$ is a weight that depends on the proposed treatment model.  Note that in the linear mean-model case with no treatment effect modification, IPW methods recover marginal parameters such as the ATE which coincide with conditional parameters such as those that appear in \eqref{eq:RMN1}, but this correspondence between marginal and conditional parameters does not follow in more general models.

\subsection{Doubly robust procedures for non-linear models}
\label{sec:dr}

Doubly robust procedures provide correct inference even when one of the component models is mis-specified. In the linear case, \eqref{eq:RMN2} yields a doubly robust procedure provided the treatment effect model is correctly specified.  The same conclusion follows for the inverse probability weighting method of section \ref{sec:ipw}, even under a slight relaxation of assumptions concerning the treatment effect model: if the propensity model is correctly specified, the ATE can be correctly estimated even if the treatment effect model is mis-specified.  Beyond the linear case, the situation is more complicated: in the log-linear equivalent to \eqref{eq:RMN1} or \eqref{eq:RMN2}, parameters in the conditional model are not equivalent to marginal parameters. One important complication is that ensuring double robustness is not as straightforward.

Consider the data generating model based on a Poisson assumption, so that conditionally the outcome $Y$ is Poisson distributed with $\log \E[Y|X,Z] = X_{0} \xi + Z \psi$.  Here $\psi$ captures the effect of treatment in the conditional model, but is not itself the ATE.  The ATE can be measured on the additive scale, as in the linear case, $\E[Y(1)] - \E[Y(0)] = \E[\exp\{X_{0} \xi + \psi \}] - \E[\exp\{X_{0} \xi\}]$, or on the multiplicative scale, for example $\E[Y(1)]/\E[Y(0)]  = \E[\exp\{X_{0} \xi + \psi \}]/\E[\exp\{X_{0} \xi\}] = \exp\{\psi\}$.

However, if the fitted Poisson regression model is mis-specified in the treatment-free component, say $\log \E[Y|X,Z] = X \beta + Z \psi$, parameter $\psi$, and hence the ATE, cannot be recovered using a standard parametric analysis.  The log-likelihood derived from this mis-specified outcome model with score function will lead to inconsistent inference for $\psi$, and the posterior distribution will concentrate at the wrong location.  Unlike in the linear case, this cannot be rectified by the inclusion of the fitted propensity score in the mean model.  The solution to this problem, first proposed by \cite{RobinsMarkNewey1992}, is to modify the likelihood-based score equation to become
\begin{equation}\label{eq:PoissonDREstEq}
\sum_{i=1}^n \begin{pmatrix} x_i^\top \\ z_i - b(x_i;\widehat \gamma) \end{pmatrix} \exp\{ - z_i \psi\} (y_i - \exp\{x_i \beta + z_i \psi\}) = 0
\end{equation}
where $b(x_i;\widehat \gamma)$ is the fitted propensity score, which can be shown to be a doubly robust estimating equation.

It is important to note that \textbf{there is no likelihood model} that corresponds to the estimating equation in \eqref{eq:PoissonDREstEq}, and consequently, no conventional Bayesian analysis that can be carried out in a doubly robust fashion.  However, the methods outlined in section \ref{sec:BBootsec} and based on the Dirichlet process/Bayesian bootstrap can be implemented, using the connection to estimating equations described in section \ref{sec:EstEquation}, with the derivative of the utility/loss function chosen to match the form in \eqref{eq:PoissonDREstEq}, with computation of the posterior samples following \eqref{eq:EstEqDP}.  It should be noted that IPW methods following the ideas in section \ref{sec:ipw} can also be used to estimate the ATE.

\subsection{Average treatment effect on the treated}
	
\label{sec:att}

In the binary treatment case, it is sometimes required to estimate the \textit{average treatment effect on the treated} (ATT), that is, the causal effect of treatment on the subgroup of individuals in the sampled population who actually received treatment.  Using counterfactual notation, the ATT is defined as the difference $\E[Y(1) - Y(0) | Z = 1]$.  Using conventional random variable notation, it is less straightforward to define this quantity, which would be problematic for conventional Bayesian analysis.  However, we may posit a new binary random variable $V$ that is assigned independently of $X$ given $Z$; $V$ can be considered a re-randomization indicator used to define two hypothetical subgroups of the treated group.  We can write the ATT as $\E[Y|X,V=1,Z=1] - \E[Y|X,V=0,Z=1]$ and use this to define an estimator based on a weighting procedure.  Crucially, the variable $V$ does not need to be observed for inference, and we can estimate the ATT from the observed data; in the simplest formulation, $\E[Y(1)| Z = 1]$ is estimated directly from the treated individuals, but $\E[Y(0) | Z = 1]$ is estimated from the untreated individuals reweighted by a case weight $w(X) = b(X)/(1-b(X))$.  Extension to a doubly robust estimator is straightforward by augmentation.

The Bayesian bootstrap procedure from section \ref{sec:BBootsec} can be used to compute a fully Bayesian posterior distribution for the ATT by using the utility function
\[
u_\theta(o,\vartheta) = - w(x,z;\vartheta_2^\opt) \log f_1(y|x,z;\vartheta_1) - \log f_2(z|x;\vartheta_2)
\]
where (see \citet{Moodie2018ATT}) the weighting function is given by
\[
w(x,z;\vartheta_2^\opt) = z + (1-z) \frac{f_2(1|x;\vartheta_2^\opt)}{f_2(0|x;\vartheta_2^\opt)} = z + (1-z) \frac{b(x;\vartheta_2^\opt)}{1-b(x;\vartheta_2^\opt)}.
\]
As for other weighting settings, it is not straightforward to estimate the ATT by simply modelling the dependence of $Y$ on $X$ and $Z$ in an outcome regression model without relying on an assumption of correct specification.

\subsection{Multiple treatments and the marginal structural model}

\label{sec:msm}

If causal inference is required for multiple treatments, then there are data generating mechanisms for which the causal effect \textbf{cannot} be inferred by modelling the outcome as a function of the treatments and confounders, no matter how complex this model is.  For a simple illustration, consider two binary treatments $(Z_1,Z_2)$ generated by the structural data generating model with $X_1 \sim Normal(1,1)$, $Z_1 \sim Bernoulli(\text{expit}(-2 + X_1))$ at the first stage, and $X_2 \sim Normal(-3+X_1+Z_1,1)$ and $Z_2 \sim Bernoulli(\text{expit}(2 - X_2))$ at the second stage, with outcome model $Y \sim Normal(X_1+Z_1+X_2+Z_2,1)$.  In this model, intervening to set $(Z_1,Z_2) = (z_1,z_2)$ yields the expected (counterfactual) outcome
\begin{align*}
  \E[Y(z_1,z_2)] & = \E_{X_1,X_2}[X_1+z_1+X_2+z_2]  = 1 + z_1 + z_2 + \E_{X_1}[\E_{X_2|X_1}[X_2|X_1]]\\[6pt]
  & = 1 + z_1 + z_2 +  \E_{X_1}[-3+X_1+z_1] = -1 + 2 z_1 + z_2.
\end{align*}
That is, $\E[Y(0,0)] = -1, \E[Y(1,0)] = 1, \E[Y(0,1)] = 0, \E[Y(1,1)] = 2$.  A correctly specified outcome model, however, consistently estimates the coefficients of $(X_1,Z_1,X_2,Z_2)$ as $(1,1,1,1)$ via ordinary least squares (or any standard Bayesian method), and therefore the counterfactual outcomes are inconsistently estimated if the standard plug-in type approach is used.  The issue arises due to the confounding that is present in the data generating model, but also due to mediation of the effect of $Z_1$ through $X_2$.

The inverse weighting approach provides a solution to this problem; with utility
\[
u_\theta(o,\vartheta) = - w(x,z;\vartheta_2^\opt) \log f_1(y|x_1,z_1,x_2,z_2;\vartheta_1) - \log f_{2}(z_1,z_2|x_1,x_2;\vartheta_2)
\]
where $w(x,z;\vartheta_2^\opt) = 1/\{f_{21}(z_1|x_1;\vartheta_2^\opt) f_{22}(z_2|x_1,z_1,x_2;\vartheta_2^\opt)\}$, and if the stagewise treatment models $f_{21}(z_1|x_1;\vartheta_2)$ and $f_{22}(z_2|x_1,z_1,x_2;\vartheta_2)$ are correctly specified, then the counterfactual quantities, and the associated average treatment effects, can be correctly inferred using the method of section \ref{sec:BBootsec}; see \cite{saarela2015bayesian}.

\section{Summary}\label{conc}

When causal inference is the aim of a statistical analysis, control of confounding is an essential consideration. If an outcome model can be correctly specified or flexibly approximated, causal inferences may follow with or without the use of propensity score methods. However, when it is not possible to correctly capture the outcome process, propensity score methods can be very valuable, particularly when the treatment allocation process is easier to characterize.  A joint modelling approach to the estimation of the propensity score and outcome model parameters can result in feedback from the outcome into the propensity score which prevents the estimated propensity score from providing balance, thus resulting in biased estimators of the treatment effect. Techniques aimed at cutting feedback have been suggested; we recap the reasoning as to why a Bayesian two-step approach, rather than one that cuts feedback is the correct approach to pursue, even if in large samples, a cutting feedback approach can provide adequate results.  We demonstrated that the standard Bayesian two-step estimator results in poor frequentist performance, but shown that this can be rectified by using the Bayesian bootstrap with linkage between the two component models, yielding a fully Bayesian procedure with good frequentist properties.

Our argument is based on the realization that the causal analysis is carried out under conscious mis-specification of the Bayesian model, and develop the framework reflecting the literature on Bayesian analysis under mis-specification \citep{walker2013} in the causal problem.   The causal setting gives a concrete example where inference under a mis-specified model -- that is, where the target of inference is not a parameter in the data generating model -- is actually the objective.   Methods that posit the capability of recovering the correct components of the outcome model using flexible modelling without reference to the propensity score also provide valid routes to inference about this target, but these methods often carry a heavier computational burden.  There are also links to modularized Bayesian inference \citep{bayarri2009,jacob2017better} which also depend on a `conscious mis-specification' formulation, and in the causal setting (the main examples and the examples in section \ref{sec:Extensions}) existing frequentist semiparametric theory can give insight into the operating characteristics of such Bayesian analyses; see \cite{pompe2021asymptotics} for initial explorations in this direction.

The Bayesian bootstrap described in section \ref{sec:BBootsec} relies on the limiting Dirichlet process specification with $\alpha \longrightarrow 0$, although equation \eqref{eq:PostPredEstLogMisspecSampDP} indicates that a more general model with $\alpha > 0$ can be deployed.  In the inference methodology described in section \ref{sec:Bims}, the requirement is simply to be able to sample independently from the posterior predictive distribution, where that distribution is consistent for the data generating process; this can be achieved by statistical procedures beyond those based on the Dirichlet process.

In this paper, we have not discussed propensity score matching methods in detail.  Such methods have been deployed successfully \citep{Liao2020} by using the propensity score to create a matched sample of treated and untreated individuals.  The principles outlined in this paper suggest that matching on an estimated propensity score, rather than averaging over the posterior distribution of the propensity score parameters, would provide superior inference, although this would arguably depend on the matching criterion used.  This is an interesting direction for future research.

\section*{Acknowledgments}

DAS, EEMM, and ASM are all supported by individual Discovery Grants from the Natural Sciences and Engineering Research Council of Canada (NSERC). WSN was supported by awards from the Conselho Nacional de Desenvolvimento Científico e Tecnológico (CNPq), Brazil (Scholarship 140529/2017-9), and Fundação de Amparo à Pesquisa do Estado do Rio de Janeiro (FAPERJ), Brazil (Scholarship E-26/200.809/2019). EEMM is a Canada Research Chair and holds a career award from the Fonds de recherche du Qu\'ebec - Sant\'e. WSN was also funded by the Emerging Leaders in the Americas Program, with the support of the Government of Canada.

	\bibliographystyle{chicago}
	\bibliography{bib/BibReferencesUpdate}

\FloatBarrier
\newpage

	\appendix
\pagenumbering{arabic}
\setcounter{page}{1}
\setcounter{table}{0}
\renewcommand{\thetable}{\Alph{section}\arabic{table}}
\setcounter{figure}{0}
\renewcommand{\thefigure}{\Alph{section}\arabic{figure}}

\section{Balance via the propensity score}
\label{sec:AppPS}

For two confounders $X_1$ and $X_2$, suppose that the propensity score is defined by
\[
\Pr[Z=1|X_1,X_2] = \text{expit}(X_1+X_2),
\]
so that the conditional probability that $Z=1$ is entirely determined by the sum $X_1+X_2$.  We may consider a parametric model $b(x;\gamma) = \text{expit}(\gamma_{1} x_1 + \gamma_{2} x_2)$, so that $\gamma_0 = (\gamma_{01},\gamma_{02})^\top = (1,1)^\top$. If $B_0  = b(X;\gamma_0)$, then \[
\Pr[Z=1|X_1,X_2,B_0]  =  B_0 \equiv \Pr[Z=1|B_0]
\]
as required.  If $B_\gamma =  b(X;\gamma) = \text{expit}(\gamma_1 X_1 + \gamma_2 X_2)$ for $\gamma \neq \gamma_0$, then $B_\gamma \neq B_0$, but
\[
P[Z = 1 | X_1, X_2, B_\gamma] = B_0 \neq B_\gamma = P[Z = 1 | B_\gamma]
\]
and we still need to know the values of $X_1$ and $X_2$ (or the value of $X_1+X_2$, or $B_0$ itself) to compute the required treatment probability; knowledge of $\gamma_1 X_1 + \gamma_2 X_2$ alone is not sufficient.  Thus we do not obtain the required conditional independence of $Z$ and $(X_1,X_2)$ after conditioning on $B_\gamma$, and $B_\gamma$ is not a balancing score.

\section{Propensity score regression in the Normal model}
\subsection{Bias and variance}
\label{sec:AppBias}
If the treatments are conditionally Normally distributed, then identical logic applies in the balancing argument (see for example \cite{imaivandyk2004}), and we may use the (fitted) conditional mean in a linear regression model for $Z$ as the balancing score.  We study the Normal case here due to its analytic tractability. Suppose that
\begin{equation}\label{eq:NE11}
Y_i = \beta_0 + \beta_1 X_{i1} + \beta_2 X_{i2} + \beta_3 X_{i3} + \tau Z_i + \epsilon_i
\end{equation}
with $\tau=5$ and $(\beta_0,\beta_1,\beta_2,\beta_3) = (3,-2,10,6)$, with $\epsilon_i \sim Normal(0,\sigma_Y^2)$ with $\sigma_Y= 1$, and suppose $Z_i \sim Normal(X_{i} \gamma_0,5^2)$, where $X_{i} = (X_{i1},X_{i2},X_{i3})$ and $\gamma_0 = (5,5,-3,2)^\top$.  The propensity score regression model is implemented by first fitting a model for $Z$ given $X$, obtaining the predicted values $\widehat b_i =  \widehat\gamma_{0} + \widehat\gamma_{1} x_{i1} + \widehat\gamma_{2} x_{i2} + \widehat \gamma_{3} x_{i3}$, and then fitting the model
\begin{equation}\label{eq:PSR00}
\E[Y|X=x,Z=z,B=b;\beta,\phi,\tau] = \beta_0 + \phi b+ \tau z
\end{equation}
which is mispecified in its treatment-free component, but correctly specified in terms of the treatment-effect component.  Confounders are simulated with mean $(2,-1,0.5)^\top$ with $\textrm{Cov}(X_j,X_k) = 0.8^{|j-k|}$ for $j,k=1,2,3$.

This model is analytically tractable and the Bayesian posterior mean, used to estimate $\tau$, can be computed for the following four models: (i) an \textit{unadjusted} analysis, where a simple linear regression assuming the conditional mean is $\beta_0 + \tau z$ is used; (ii) a propensity score regression (PSR) model of the form of \eqref{eq:PSR0} using the true propensity score values; (iii) a PSR model using the estimated propensity score in a two-step analysis; (iv) a PSR model using the estimated propensity score in a cut feedback analysis.

\begin{figure}[ht]
  \centering
  \includegraphics[width=\textwidth]{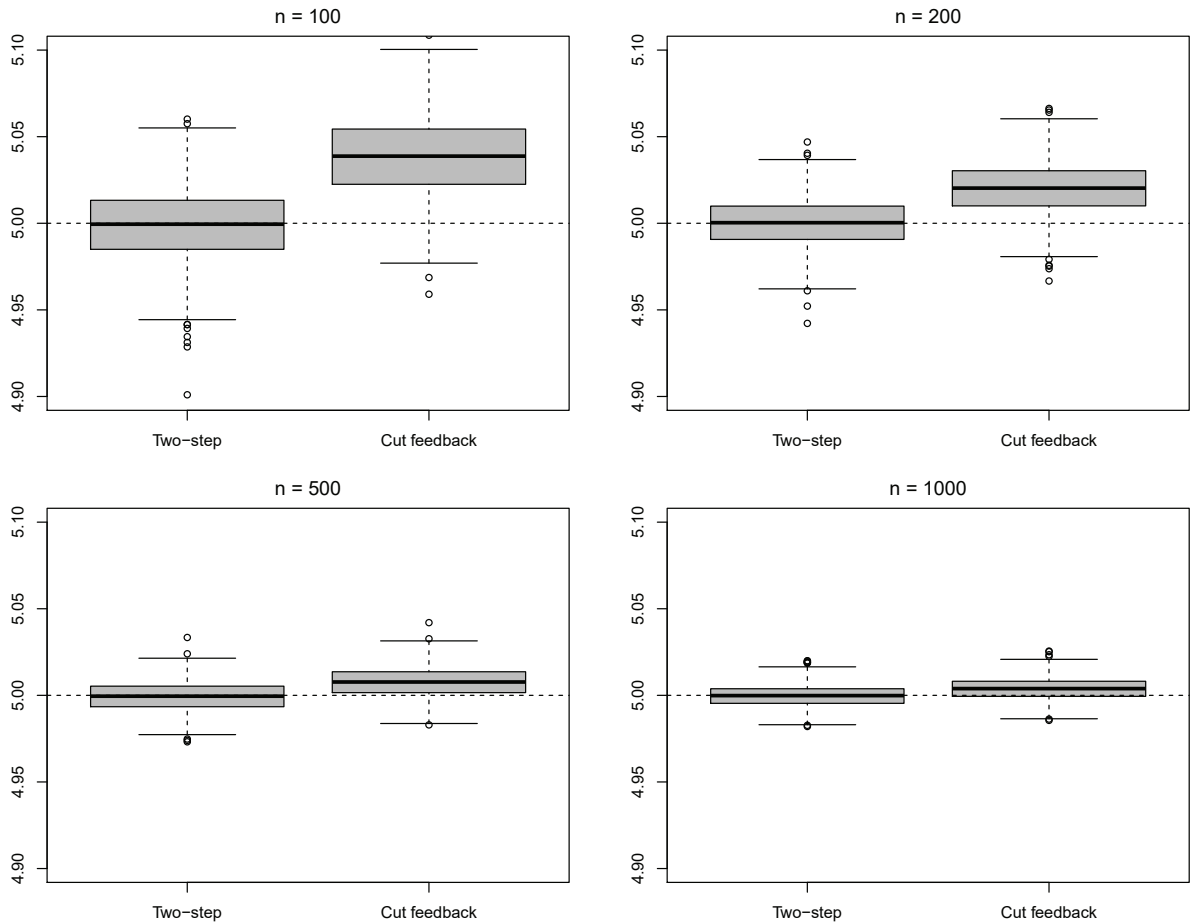}
  \caption{Boxplots of Bayesian posterior mean estimates of $\tau$ using two-step or cut feedback procedures.  1000 replicates for sample sizes $n=100,200,500,1000$}\label{fig:sim1box}
\end{figure}

A simulation study of 1000 replicate analyses illustrates the bias, standard deviation and root mean square error (RMSE) of the Bayesian estimates derived from four sample sizes $n=100,200,500,1000$; the $n=100$ case is studied here and not in the other simulation studies as it highlights the differences in results more concretely.  Table \ref{tab:sim1tab} contains the numerical values for these quantities for four different sample sizes, whereas Figure \ref{fig:sim1box} depicts the boxplots for the two-step and cut feedback analyses.  These results show that both the propensity score method using the true propensity score model and the two-step approach yield unbiased procedures, whereas the cut feedback approach produces bias and higher RMSE.  Bias, standard deviation and RMSE decrease as the sample size increases. The bias of the cut feedback method can be mitigated by the use of a more complex treatment-free model, although the bias is not removed.

\subsubsection{Estimation using the true propensity score}

Note that use of the true propensity score results in a \textbf{larger} RMSE than when the propensity score is estimated: this is an example of a phenomenon that can occur in situations where plug-in methods are used in estimating equations that deviate from `full likelihood'-based estimation procedures. Essentially, in the frequentist calculation, the phenomenon arises (i) when the two sets of parameters in outcome and treatment models are estimated using separate estimating functions, with estimates from the latter plugged into the former, and (ii) the projection of the estimating function for the outcome parameters onto the space spanned by the estimating function for the treatment parameters has a smaller norm than the estimating function that assumes the treatment parameters to be known.   A sufficient condition for it to transpire is the asymptotic independence of estimators originating from the two component models.  The result was established explicitly for the two-step propensity score regression model as in \eqref{eq:PSR00} by \cite{henmieguchi2004}; see also \citet{Pierce1982,RobinsMarkNewey1992}.  \begin{table}[ht]
  \centering
  \caption{Simulation for Normal example: bias, standard deviation, root mean square error (RMSE) for 1000 replicate data sets for sample sizes $n=100,200,500,1000$ for each of four methods. }\label{tab:sim1tab}
  \smallskip
\begin{tabular}{|l|ccc|ccc|}
  \hline
         & \multicolumn{3}{c|}{Unadjusted} &  \multicolumn{3}{c|}{True $\gamma_0$} \\
\multicolumn{1}{|c|}{$n$}& Bias & s.d. & RMSE & Bias & s.d. & RMSE \\
         \hline
$100$  & 0.8282 &0.1949 &0.8508 &0.0007 &0.2116 &0.2115 \\
$200$  & 0.8328 &0.1372 &0.8440 &0.0020 &0.1567 &0.1566 \\
$500$  & 0.8282 &0.0857 &0.8326 &0.0031 &0.0944 &0.0945 \\
$1000$ & 0.8311 &0.0621 &0.8334 &0.0005 &0.0665 &0.0665 \\
  \hline
         & \multicolumn{3}{c|}{Two-step} & \multicolumn{3}{c|}{Cut feedback}\\
\multicolumn{1}{|c|}{$n$}& Bias & s.d. & RMSE & Bias & s.d. & RMSE \\
         \hline
$100$  & -0.0007 &0.0214 &0.0214 &0.0383 &0.0234 &0.0449\\
$200$  & ~0.0002 &0.0144 &0.0144 &0.0200 &0.0150 &0.0250\\
$500$  & -0.0005 &0.0088 &0.0088 &0.0076 &0.0090 &0.0117\\
$1000$ & -0.0001 &0.0062 &0.0062 &0.0039 &0.0064 &0.0075\\
  \hline
\end{tabular}
\end{table}

\subsection{Undercoverage of the conventional plug-in approach} \label{sec:undercover}
Consider the following data generating mechanism with Normal outcome and binary treatment models. Suppose the outcome model is specified as
\begin{equation}\label{eq:NE1}
Y_i = \beta_0 + \beta_1 X_{i1} + \beta_2 X_{i2} + \beta_3 X_{i3} + \tau Z_i + \epsilon_i
\end{equation}
with $\tau=5$ and $(\beta_0,\beta_1,\beta_2,\beta_3) = (3,-2,10,6)$, and $\epsilon_i \sim Normal(0,1)$.  In the treatment assignment model, suppose that we have $Z_i |X_i = x_i; \gamma_0 \sim Bernoulli(p_i)$, with $\text{logit} (p_i) = \gamma_{00} + \gamma_{01} x_{i01} + \gamma_{02} x_{i2} + \gamma_{03} x_{i3}$ for $ \gamma_0 = (2,-2,-2,1)^\top$. Confounders are simulated with mean $(2,-1,0.5)^\top$ with $\textrm{Cov}(X_j,X_k) = 0.8^{|j-k|}$ for $j,k=1,2,3$.

The propensity score regression model is implemented by first fitting a Bayesian model for $Z$ given $X$, obtaining the predicted values $\widehat b_i =  \widehat\gamma_{0} + \widehat\gamma_{1} x_{i1} + \widehat\gamma_{2} x_{i2} + \widehat \gamma_{3} x_{i3}$, and then fitting the regression model
\begin{equation}\label{eq:PSR0}
\E[Y|X=x,Z=z,B=\widehat b;\beta,\phi,\tau] = \beta_0 + \phi \widehat b+ \tau z
\end{equation}
which is mis-specified in its treatment-free component, but correctly specified in terms of the treatment-effect component.  With a flat prior for $(\phi,\tau)$ the posterior distribution is bivariate Normal, and the marginal posterior for $\tau$ is univariate normal.  From this calculation, a 95\% credible interval for $\tau$ can be constructed.  Note that the credible interval is not primarily motivated by notions of frequentist (coverage) properties.  However, in an ordinary Normal linear regression analysis, coverage of a 95\% Bayesian interval would be at the nominal level at least in large samples.

In 2000 replicate data sets, and for four sample sizes, we compare the coverage of a Bayesian interval for $\tau$ arising from a correctly specified model \eqref{eq:NE1} with the coverage of an interval from the propensity score regression model \eqref{eq:PSR0}.  The results are contained in the first two panels of Table \ref{tab:coverage}.  The frequentist bias of the Bayesian estimator is zero for both methods, and as expected the variance of the estimator in the correctly specified model is smaller than that for the PSR model.  However, whereas the coverage of the Bayesian interval in the correctly specified model is at the nominal level, the coverage for the PSR model is below the nominal level even for large $n$.

\begin{table}[ht]
\setlength{\tabcolsep}{2pt}
  \centering
\caption{Frequentist properties of Bayesian estimators: $\sqrt{n}$ times the standard deviation, and coverage (Cov.) of 95\% interval, in 2000 replicate samples using the exact regression model (Exact), a two-step propensity score regression model (PSR), a PSR with frequentist bootstrap, and a PSR with Bayesian bootstrap (section \ref{sec:BBootsec}). \label{tab:coverage}}
\begin{tabular}{|r|cc|cc|cc|cc|}
\multicolumn{9}{c}{}\\[-6pt]
  \hline
 \multicolumn{1}{|c|}{$n$}& \multicolumn{2}{c|}{Exact} & \multicolumn{2}{c|}{PSR}& \multicolumn{2}{c|}{Boot PSR} & \multicolumn{2}{c|}{Bayesian Boot.} \\
 & $\sqrt{n} \times \textrm{s.d.}$ & Cov. & $\sqrt{n} \times \textrm{s.d.}$ & Cov.& $\sqrt{n} \times \textrm{s.d.}$ & Cov.  & $\sqrt{n} \times \textrm{s.d.}$ & Cov. \\
\hline
$200$   & 2.623 & 95.12 & 4.075 & 81.64& 3.924  &  95.60 & 3.958  &  94.30\\
$500$   & 2.589 & 94.92 & 4.032 & 81.27& 3.955  &  94.60 & 3.913  &  94.10\\
$1000$  & 2.569 & 95.38 & 3.985 & 81.34& 3.974  &  94.60 & 3.890  &  94.75\\
$2000$  & 2.589 & 95.35 & 3.981 & 81.27& 3.929  &  94.65 & 3.925  &  94.65\\
  \hline
\end{tabular}
\end{table}

The undercoverage phenomenon arises because of the lack of acknowledgement of the plug-in estimation for $\gamma$, and the fact that in replicate data sets entire triples of $(x,y,z)$ variables are generated.  Coverage is an assessment of the expected behaviour of the credible interval under the true data generating distribution, but the model \eqref{eq:PSR0} is mis-specified and does not match the data generating distribution.  The conventional Bayesian two-step plug-in method therefore does not have reasonable frequency properties. To obtain better coverage, we must revert to the Bayesian inference approach for mis-specified models described in section \ref{sec:Bims}, and deploy the Bayesian bootstrap approach from section \ref{sec:BBootsec}.  The fourth column of Table \ref{tab:coverage} displays the results for the Bayesian bootstrap analysis of the propensity score regression model, whereas the third column contains the results for the frequentist non-parametric bootstrap.  The non-parametric model underpinning the Bayesian bootstrap overcomes issues of undercoverage.


\section{Further comparison with Bayesian causal forests}
\label{sec:AppBCF}
\begin{figure}[!bt]
	\centering
    \vspace{-0.5 in}
	\includegraphics[scale=.45]{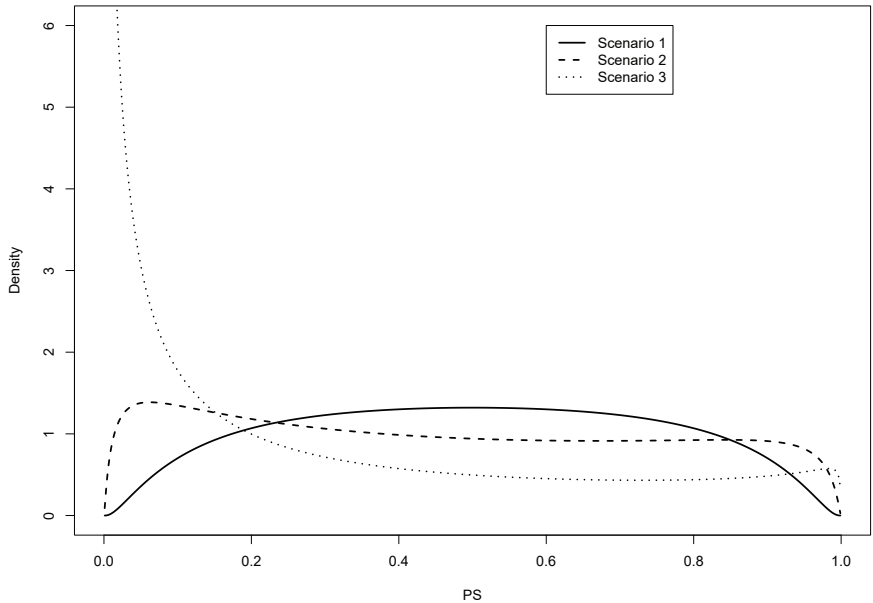}
	\caption{Simulated example 2: Scenarios considered for the exposure model in the true data generating mechanism with binary exposure.}
	\label{fig:ps}
\end{figure}

\FloatBarrier
We use the same simulation set up as for Example 2.  We generate $p = 4$ confounders, $X=(X_1,X_2,X_3,X_4)$ independently, with $X_1,X_2 \sim Normal(1,1)$, and $X_3, X_4 \sim Normal(-1,1)$.  For the treatment assignment we simulate $Z_i \sim Bernoulli(p_i)$ with
\begin{equation*}
\mbox{logit} (p_i) = \gamma_0 + \gamma_1 x_{i1} + \gamma_2 x_{i2} + \gamma_3x_{i3} + \gamma_4x_{i4},
\end{equation*}
and three settings of the parameters:
\begin{itemize}

\item Scenario 1: $\gamma = (0.0,0.3,0.8,0.3,0.8)$;

\item Scenario 2: $\gamma = (0.5,0.5,0.75,1.0,1.0)$;

\item Scenario 3: $\gamma = (0.0,0.45,0.90,1.35,1.8)$.

\end{itemize}
The propensity score distributions are displayed in Figure \ref{fig:ps}.

We study two outcome model cases: in each case we simulate $Y \sim Normal(\mu_Y,1)$ and change the $\mu_Y$ specification in the data generating process.

\begin{enumerate}[1.]

\item \textit{\textbf{Treatment effect with modification:}} For the outcome model, we simulate $Y$ with
\[
\mu_Y = 0.25x_{1} + 0.25x_{2} + 0.25x_{3} + 0.25x_{4} + 1.5x_{3}x_{4} + (\psi_0 + \psi_1 x_1) z
\]
In the two-step method, the fitted mean model takes the form
\[
\beta_0 + (\psi_0 + \psi_1 x_1) z + (\phi_0 + \phi_1 x_1) \widehat b
\]
where $\widehat b$ is obtained from the fit of the correctly specified propensity model.  For the BCF analysis we specify
\[
\mu(x,\widehat{b}(x)) + \tau(x_1,\widehat{b}(x)) z.
\]
where $x=(x_1,x_2,x_3,x_4)$.

\item \textit{\textbf{No treatment effect: }}For the outcome model, we simulate $Y$ with no treatment effect
\[
\mu_Y = 0.25x_{1} + 0.25x_{2} + 0.25x_{3} + 0.25x_{4} + 1.5x_{3}x_{4}
\]
In the two-step method, the fitted mean model takes the form
\[
\beta_0 + \tau z + \phi_0 \widehat b
\]
where $\widehat b$ is obtained from the fit of the correctly specified propensity model.  For the BCF analysis we specify
\[
\mu(x,\widehat{b}(x)) + \tau(\widehat{b}(x)) z.
\]
where $x=(x_1,x_2,x_3,x_4)$.

\end{enumerate}

In the two-step analysis, we assign $Normal(0,10^2)$ priors for the elements of $\gamma$, but non-informative priors for the parameters in the outcome model.  We compare the results for the two-step method and the Bayesian Causal Forests (BCF) method for sample sizes $n=200,500,1000$ and $2000$.

\begin{table}[ht]
\centering
	\caption{Additional simulated example: Comparison of results for two-step fitted using the Linked Bayesian bootstrap, and the BCF method for different propensity score settings. Summary of 2000 Bayesian estimates and credible intervals.  Rows correspond to the bias, RMSE, and coverage rates for Scenario 1, Scenario 2 and Scenario 3.   \label{tab:Ex3extendedNonNull}}
\begin{tabular}{ccl|rrrr}
&& & \multicolumn{4}{c}{$n$}\\
&& Method & 200 & 500 & 1000 & 2000 \\
  \hline
\multirow{6}{*}{\rotatebox[origin=c]{90}{Scenario 1}}&
\multirow{2}{*}{Bias}&2S & -0.004 & 0.004 & 0.004 & 0.001 \\
  &&BCF & 0.058 & 0.028 & 0.019 & 0.012 \\
&\multirow{2}{*}{RMSE}&
  2S & 0.308 & 0.200 & 0.140 & 0.098 \\
  &&BCF & 0.277 & 0.161 & 0.110 & 0.076 \\
&\multirow{2}{*}{Coverage}&
  2S & 93.2 & 93.3 & 93.4 & 94.3 \\
  &&BCF & 90.8 & 88.3 & 87.2 & 86.1 \\
  \hline
\multirow{6}{*}{\rotatebox[origin=c]{90}{Scenario 2}}&
\multirow{2}{*}{Bias}&  2S & -0.019 & -0.003 & -0.003 & 0.001 \\
  &&BCF & 0.114 & 0.068 & 0.043 & 0.03 \\
&\multirow{2}{*}{RMSE}&
  2S & 0.308 & 0.192 & 0.138 & 0.099 \\
  &&BCF & 0.318 & 0.184 & 0.121 & 0.087 \\
&\multirow{2}{*}{Coverage}&
  2S & 94.0 & 93.8 & 94.0 & 94.1 \\
  &&BCF & 91.4 & 89.6 & 89.2 & 88.1 \\
\hline
\multirow{6}{*}{\rotatebox[origin=c]{90}{Scenario 2}}&
\multirow{2}{*}{Bias}&  2S & -0.017 & -0.009 & -0.001 & 0.001 \\
  &&BCF & 0.209 & 0.099 & 0.072 & 0.047 \\
&\multirow{2}{*}{RMSE}&
  2S & 0.356 & 0.215 & 0.152 & 0.110 \\
  &&BCF & 0.435 & 0.247 & 0.165 & 0.115 \\
&\multirow{2}{*}{Coverage}&
  2S & 93.3 & 93.6 & 94.7 & 94.8 \\
  &&BCF & 93.4 & 93.0 & 92.0 & 90.4 \\
   \hline
\end{tabular}
\end{table}

\begin{table}[ht]
\centering
	\caption{Additional simulated example: (no treatment effect): Comparison of results for two-step fitted using the Linked Bayesian bootstrap, and the BCF method for different propensity score settings. Summary of 2000 Bayesian estimates and credible intervals.  Rows correspond to the bias, RMSE, and coverage rates for Scenario 1, Scenario 2 and Scenario 3.   \label{tab:Ex3extendedNull}}
\begin{tabular}{ccl|rrrr}
&& & \multicolumn{4}{c}{$n$}\\
&& Method & 200 & 500 & 1000 & 2000 \\
  \hline
\multirow{6}{*}{\rotatebox[origin=c]{90}{Scenario 1}}&
\multirow{2}{*}{Bias}&
  2S & -0.003 & 0.006 & 0.005 & 0.001 \\
  &&BCF & 0.019 & 0.009 & 0.007 & 0.004 \\
&\multirow{2}{*}{RMSE}&
  2S & 0.265 & 0.173 & 0.119 & 0.084 \\
  &&BCF & 0.167 & 0.102 & 0.067 & 0.046 \\
&\multirow{2}{*}{Coverage}&
  2S & 93.0 & 93.1 & 93.3 & 94.6 \\
&&BCF & 97.7 & 96.9 & 97.4 & 97.4 \\
  \hline
\multirow{6}{*}{\rotatebox[origin=c]{90}{Scenario 2}}&
\multirow{2}{*}{Bias}&
  2S & -0.012 & -0.000 & -0.001 & 0.001 \\
  &&BCF & 0.028 & 0.018 & 0.013 & 0.012 \\
&\multirow{2}{*}{RMSE}&
  2S & 0.264 & 0.167 & 0.119 & 0.085 \\
  &&BCF & 0.179 & 0.105 & 0.073 & 0.052 \\
&\multirow{2}{*}{Coverage}&
  2S & 92.8 & 93.2 & 94.4 & 93.6 \\
  &&BCF & 98.3 & 97.6 & 97.3 & 96.2 \\
  \hline
\multirow{6}{*}{\rotatebox[origin=c]{90}{Scenario 3}}&
\multirow{2}{*}{Bias}&
  2S & 0.008 & 0.001 & -0.001 & 0.001 \\
  &&BCF & 0.059 & 0.019 & 0.011 & 0.006 \\
&\multirow{2}{*}{RMSE}&
  2S & 0.316 & 0.192 & 0.138 & 0.100 \\
  &&BCF & 0.228 & 0.138 & 0.096 & 0.069 \\
&\multirow{2}{*}{Coverage}&
  2S & 92.3 & 94.0 & 94.0 & 94.4 \\
  &&BCF & 98.8 & 98.2 & 97.8 & 96.6 \\
   \hline
\end{tabular}
\end{table}

\end{document}